\documentclass{LMCS}

\def\doi{8(3:22)2012}
\lmcsheading%
{\doi}
{1--31}
{}
{}
{Jul.~\phantom06, 2011}
{Sep.~21, 2012}
{}

\usepackage{enumerate}
\usepackage{hyperref}

\usepackage{stmaryrd}
\usepackage{proof}
\usepackage{amsmath}
\usepackage{amssymb}
\usepackage[all]{xy}

\newcommand{\blackslug}{\rule{7pt}{7pt}}
\renewcommand{\vec}[1]{\mathbf{#1}}

\newcommand{\wrong}{\mathit{err}}
\newcommand{\diag}{\Delta}
\newcommand{\defeq}{\ensuremath{\stackrel{\mbox{\tiny \it def}}{=}}}
\newcommand{\defiff}{\ensuremath{\stackrel{\mbox{\tiny \it def}}{\iff}}}
\newcommand{\sem}[1]{[\![{#1}]\!]}
\newcommand{\semprim}[1]{\llparenthesis{#1}\rrparenthesis}
\newcommand{\semun}[1]{[\![{#1}]\!]^1}
\newcommand{\semn}[1]{[\![{#1}]\!]^n}
\newcommand{\sembin}[1]{[\![{#1}]\!]^2}

\newcommand{\dom}{\operatorname{dom}}
\newcommand{\Seg}{\mathbf{S}}

\newcommand{\avar}{\mathsf{AVar}}
\newcommand{\var}{\mathsf{Var}}
\newcommand{\heap}{\mathsf{Heap}}
\newcommand{\irel}{\mathsf{IRel}}
\newcommand{\pinteger}{\mathsf{PosInt}}
\newcommand{\integer}{\mathsf{Int}}
\newcommand{\nat}{\mathsf{Nat}}
\newcommand{\fin}{{\mbox{\tiny \it fin}}}
\newcommand{\FV}{\mathsf{FV}}

\newcommand{\ipsto}{{\hookrightarrow}}

\newcommand{\blank}{{\_}}
\newcommand{\true}{\mathsf{true}}
\newcommand{\false}{\mathsf{false}}

\newcommand{\code}[1]{\mathtt{#1}}
\newcommand{\init}{\code{init}}
\newcommand{\gonext}{\code{nxt}}
\newcommand{\final}{\code{fin}}
\newcommand{\gfinal}{\code{fin}}
\newcommand{\bfinal}{\code{badfin}}
\newcommand{\incr}{\code{inc}}
\newcommand{\decr}{\code{dec}}
\newcommand{\cskip}{\code{skip}}

\newcommand{\klet}{\mathbf{let}}
\newcommand{\kin}{\mathbf{in}}
\newcommand{\kif}{\mathbf{if}}

\newcommand{\lcheck}{\mbox{\sc Chk}}
\newcommand{\jstar}{\mbox{jStar}}
\newcommand{\verifast}{\mbox{VeriFast}}
\newcommand{\ynot}{\mbox{Ynot}}

\newcommand{\paperskip}[1]{}
\newcommand{\proofskip}[1]{}

\begin{document}
\title{Two for the Price of One: Lifting Separation Logic Assertions}

\author[J.~Thamsborg]{Jacob Thamsborg\rsuper a}	
\address{{\lsuper{a,b}}IT University of Copenhagen, Rued Langgaards Vej 7, DK-2300 K\o{}benhavn S, Denmark}	
\email{\{thamsborg, birkedal\}@itu.dk}
\author[L.~Birkedal]{Lars Birkedal}	
\address{\vskip-6 pt}	
\author[H.~Yang]{Hongseok Yang\rsuper c}
\address{{\lsuper c}Wolfson Building, Parks Road, Oxford, OX1 3QD, UK}
\email{hongseok00@gmail.com}


\begin{abstract}
Recently, data abstraction has been studied in the context of
separation logic, with noticeable practical successes: the developed
logics have enabled clean proofs of tricky challenging programs, such 
as subject-observer patterns, and they have become the basis of
efficient verification tools for Java (jStar), C (VeriFast) and 
Hoare Type Theory (Ynot). In this paper, we give a new semantic 
analysis of such logic-based approaches using Reynolds's 
relational parametricity. The core of the analysis is our lifting theorems,
which give a sound and complete condition for 
when a true implication between assertions in the standard 
interpretation entails that the same implication holds 
in a relational interpretation. Using these theorems,
we provide an algorithm for identifying abstraction-respecting
client-side proofs; the proofs ensure that clients cannot
distinguish two appropriately-related module implementations. 
\end{abstract}


\maketitle

\section{Introduction}

Data abstraction is one of the key design principles for building
computer software, and it has been the focus of active research
from the early days of computer science.
Recently, data abstraction has been studied in the context of
separation logic \cite{Parkinson:Bierman:05,biering-birkedal-torpsmith-esop05,nanevski-htt-07,Parkinson:Bierman:08,Chin-popl08},
with noticeable practical successes: the developed
logics have enabled clean proofs of tricky challenging programs, such
as the subject-observer pattern, and they have become the basis of
efficient verification tools for Java (\jstar~\cite{DistefanoJStar08}),
C (\verifast~\cite{JacobsVeriFast08}) and
Hoare Type Theory (\ynot~\cite{BirkedalL:ynot-conf}).

In this paper, we give a new semantic
analysis of these logic-based approaches using Reynolds's
relational parametricity. Our techniques can be used to prove
\emph{representation independence}, i.e., that clients cannot distinguish
between related module implementations,
a consequence that we would expect from using data abstraction, but
(as we shall see) a consequence that only holds for certain good clients.

\begin{figure}[t]
\hrule
\vspace{2mm}
{\sc Interface Specification}
$$
\begin{array}{@{}l@{\quad}l@{\quad}l@{}}
\{1{\ipsto}\blank\}\init\{a\}
&
\{a\}\gonext\{b\}
&
\{b\}\final\{1{\ipsto}\blank\}
\\
\{a\}\incr\{a\}
&
\{b\}\decr\{b\}
\end{array}
$$
{\sc Two Implementations of a Counter}
$$
\begin{array}{@{}l@{\;\;}l@{\;\;}l@{}}
\init_1 \defeq [1]{:=}0
&
\gonext_1 \defeq \cskip
&
\final_1 \defeq \cskip
\\
\incr_1 \defeq [1]{:=}[1]{+}1
&
\decr_1 \defeq [1]{:=}[1]{-}1
\\[1.5ex]
\init_2 \defeq [1]{:=}0
&
\gonext_2 \defeq [1]{:=}{-}[1]
&
\final_2 \defeq [1]{:=}{-}[1]
\\
\incr_2 \defeq [1]{:=}[1]{+}1
&
\decr_2 \defeq [1]{:=}[1]{+}1
\end{array}
$$
{\sc Client-side Proof Attempts}
$$
\begin{array}[t]{@{}l@{}l@{}}
\{1\ipsto\blank\}\;
&
\init;\;
\{a\}
\\
&
\incr;\;
\{a\}
\\
&
\gonext;\;
\{b\}
\\
&
\decr;\;
\{b\}
\\
&
\final\;
\{1\ipsto \blank\}
\end{array}
\qquad
\begin{array}[t]{@{}l@{}l@{}}
\{1\ipsto\blank\}\;
&
\init;\;
\{a\}
\\
&
\incr;\;
\{a\}
\\
&
\gonext;\;
\{b\}
\\
&
[1]{:=}[1]{-}1;
\;
\{???\}
\end{array}
$$
\hrule
\caption{Two-stage Counter}
\label{fig:counter}
\end{figure}

\paragraph{Logic-based Data Abstraction}
The basic idea of the logic-based approaches is that the private states of
modules are exposed to clients only abstractly using \emph{assertion
variables}~\cite{biering-birkedal-torpsmith-esop05}, also known as
\emph{abstract predicates}~\cite{Parkinson:Bierman:05}\footnote{Abstract
predicates do take arguments, though. We conjecture that it is equally
expressive to use an assertion variable for each combination of abstract
predicate and concrete arguments.}. For concreteness, we consider a two-stage
counter module and client programs in Figure~\ref{fig:counter}. The module
realizes a counter with increment and decrement operations, called $\incr$
and $\decr$. An interesting feature is that the counter goes through two
stages in its lifetime; in the first stage, it can perform only the increment
operation, but in the second, it can only run the decrement. The interface
specification in the figure formalizes this intended behavior of the counter
using assertion variables $a$ and $b$, where $a$ means that the counter is in
the first stage and $b$ that the counter is in the second. The triple for
$\init$ says that the initialization can turn the assertion
$1{\ipsto}\blank$, denoting heaps with cell $1$, to the assertion variable
$a$, which describes an abstract state where we can only call $\incr$ or
$\gonext$ (since $a$ is the precondition of only those operations). The
abstract state $a$ can be changed to $b$ by calling $\gonext$, says the
triple for the $\gonext$ operation. In $b$ we are allowed to run $\decr$ but
not $\incr$. Finally, $\final$ can turn the abstract state $b$ back to
$1{\ipsto}\blank$. Note that by using $a$ and $b$, the interface
specification does not expose the private state of the module to the client.
It reveals only \emph{partial information} about the private state of the
module; here it is whether the private state is in the first or the second
stage. The flexibility afforded by revealing partial information is very
useful in applications; see the examples mentioned in the references above.

In these logic-based approaches, proof attempts for clients of a module can
succeed only when they are given with respect to the abstract interface
specification, without making any further assumptions on assertion variables.
For instance, the proof attempt on the bottom left of
Figure~\ref{fig:counter} is successful, whereas the bottom right one is not,
because the latter assumes that the assertion variable $b$ entails the
allocatedness of cell $1$. This is not so, even when the entailment holds for
an actual definition of $b$.

\paragraph{Representation Independence}
In this paper, we give a condition on client-side proofs that ensure
\emph{representation independence}: take a client with a standard proof of
correctness that satisfies this condition and two implementations of a
module; if we can relate the heap-usage of the two modules in a way preserved
by the module operations, then the client gives the same result with both
modules. To relate the heap-usage, we need to give, for each assertion
variable, a relation on heaps and verify that the module operations respect
these relations --- the \emph{coupling relations}.

As an example, consider the left-hand side client in
Figure~\ref{fig:counter}. The proof of the specification
\[
\{1\ipsto\blank\}\; \init;\; \incr;\; \gonext;\; \decr;\; \final\;
\{1\ipsto\blank\}
\]
satisfies the forthcoming condition on client-side proofs. Also, we have two
implementations of the counter module that the client makes use of; both use
cell $1$ to represent their private states, but in different ways
--- the first stores the current value of the counter, but the second stores
the current value or its negative version, depending on whether it is in the
first stage or the second. Accordingly, we give the two coupling relations:
$$
\begin{array}{@{}l@{}}
  r_a \defeq \{ (h_1,h_2) \mid 1 {\in} \dom(h_1) {\cap} \dom(h_2) \land h_1(1) {=} h_2(1)\}
\\
  r_b \defeq \{ (h_1,h_2) \mid 1 {\in} \dom(h_1) {\cap} \dom(h_2) \land h_1(1) {=} {-}h_2(1)\}
\end{array}
$$
It is easy to see that all module operations preserve these coupling
relations. If, say, $(h_1, h_2) \in r_b$, then we have $h_1(1)=n$ and
$h_2(1)=-n$ for some $n$ and so $(h_1[1 \mapsto n-1], h_2[1 \mapsto -n+1])
\in r_b$ too; hence the decrement operations of the modules respect the
coupling relation. By Theorem \ref{thm:parametricity-logic} we now get that
the client specification is valid also in a binary reading: if we take any
two heaps and run the client with one module in the first and with the other
module in the second then we will end up with two heaps holding the same
value in cell 1
--- provided that we started out with two such. Indeed, the binary
reading of the assertion $1\ipsto\blank$ is
\[
  \{ (h_1,h_2) \mid 1 {\in} \dom(h_1) \cap \dom(h_2) \land h_1(1) = h_2(1)\}
\]
which incidentally coincides with $r_a$.

It is worthwhile to emphasize that this is not a consequence of the standard
unary reading of the specification of the client: due to the existentially
quantified content of cell 1, running with the one module could yield
different contents of cell 1 than running with the other module, even if the
contents are initially the same. On the other hand, it is the presence of
this quantification that makes the binary reading worthwhile: if our client
had a more exhaustive specification, say
\[
  \{1\ipsto 0\}\; \init;\; \incr;\; \gonext;\; \decr;\; \final\; \{1\ipsto
  0\},
\]
then the standard unary reading suffices for representation
independence and the binary reading would provide no news. Often,
though, the more exhaustive specification will be harder to prove, in
particular for verification tools.

\paragraph{The Rule of Consequence and Lifting}
In earlier work~\cite{Birkedal:Yang:07} we were able to prove
such a representation independence result for a more restricted form
of logical data abstraction, namely
one given by frame rules rather than general assertion variables.
Roughly speaking, frame rules use a restricted form of assertion
variables that are not exposed to clients at all,
as can be seen from some models of separation logic in which
frame rules are modelled via quantification over semantic
assertions~\cite{Birkedal:Torp-Smith:Yang:05}. This means that
the rules do not allow the exposure of even partial information
about module internals. (On the other hand, frame
rules implement information hiding, because they
completely relieve clients of tracking the private state of a module,
even in an abstracted form.)
Our model in~\cite{Birkedal:Yang:07}
exploited this restricted use of assertion variables,
and gave relational meanings to Hoare triple specifications,
which led to representation independence.

Removing this restriction and allowing assertion variables in client proofs
turned out to be very challenging. The challenge is the use of the rule of
consequence in client-side proofs; this has implications between assertions
(possibly containing assertion variables) as hypotheses, and such do not
always \emph{lift}, i.e., they may hold in the standard, unary reading of
assertion whilst failing in the binary reading. In this paper, we provide a
sound and, in a certain sense, complete answer to when the lifting can be
done.

\paperskip{
To understand the challenges involved in more detail, consider
the example in Figure~\ref{fig:clientproof}.
Here we have a module specification and two implementations
together with two clients. We will now explain in what sense the
client on the left is a \emph{good} client and the one on the right
is a \emph{bad} client.   The client on the left calls $\init$ and ends
with a post-condition $\{1\ipsto\blank \,\wedge\, a * b\}$.
Since $\{1\ipsto\blank \,\wedge\, a * b\} \implies
\{a \vee b\}$ is true, the rule of consequence can be applied
to yield the precondition of $\gfinal$, which can be called, ending
up with postcondition $\{1\ipsto\blank\}$.
The key point here is the use of the rule of consequence.
In this case, it involves a valid implication, which can indeed be
lifted to an implication between relational interpretations
of the formulas (this is a consequence of Theorem~\ref{thm:}, see ??).
Therefore, there is a relational
interpretation of the client specification and thus we can prove
representation independence by showing that the implementations
are related, much as in the earlier example above
(see Section~\ref{sec:parametricity} for a detailed explanation).
Hence we think of this client as a \emph{good} client.
}

\begin{figure}[t]
\hrule
\vspace{2mm}
{\sc Interface Specification}
$$
\begin{array}{@{}l@{}}
\{1\ipsto\blank\}\init\{1\ipsto \blank \,\wedge\, a * b\}
\qquad
\{1\ipsto\blank\}\gfinal\{1\ipsto\blank\}
\\[0.5ex]
\{1\ipsto\blank * a \,\vee\, 1\ipsto\blank * b\}\bfinal\{1\ipsto\blank\}
\end{array}
$$
{\sc Two Implementations}
$$
\begin{array}{@{}l@{}}
\init_1 \defeq \cskip
\quad
\gfinal_1 \defeq \cskip
\quad
\bfinal_1 \defeq [1]{:=}1
\\[1ex]
\init_2 \defeq \cskip
\quad
\gfinal_2 \defeq \cskip
\quad
\bfinal_2 \defeq [1]{:=}2
\end{array}
$$
{\sc Two Client-side Proofs}
$$
\begin{array}[t]{@{}l@{}l@{}}
\{1\ipsto\blank\}\;
\\
\;\;\init;
\\
\{1\ipsto\blank \,\wedge\, a * b\}
\\
\{1\ipsto\blank\}
\\
\;\;\gfinal
\\
\{1\ipsto\blank\}
\end{array}
\qquad\qquad
\begin{array}[t]{@{}l@{}l@{}}
\{1\ipsto\blank\}
\\
\;\;\init;
\\
\{1\ipsto\blank \,\wedge\, a * b\}
\\
\{1 \ipsto\blank * a \,\vee\, 1 \ipsto\blank * b\}
\\
\;\;\bfinal
\\
\{1\ipsto\blank\}
\end{array}
$$
\hrule
\caption{Good or Bad Client-side Proofs}
\label{fig:clientproof}
\end{figure}

For instance, consider the example in Figure~\ref{fig:clientproof}. Our
results let us conclude that the client-side proof on the left is
\emph{good} but the one on the right is \emph{bad}; hence we expect to derive
representation independence only from the former. The client on the left
calls $\init$ and ends with the post-condition $(1\ipsto\blank \wedge a
* b)$. Since $(1\ipsto\blank \wedge a
* b) \implies 1\ipsto\blank$ is true in the standard interpretation, the rule
of consequence can be applied to yield the precondition of $\gfinal$, which
can be called, ending up with the postcondition $(1\ipsto\blank)$. The key
point here is the implication used in the rule of consequence. Our results
imply that this implication can indeed be lifted to an implication between
relational meanings of assertions $(1\ipsto\blank \wedge a
* b)$ and $1\ipsto\blank$ (Theorem~\ref{theorem:shadow} in Section~\ref{section:liftingtheoremsandcompleteness}).  They also
entail that this lifting implies the representation independence theorem. The
coupling relations
$$
\begin{array}{@{}l@{}}
  r_a \defeq \{ (h_1,h_2) \mid 1 {\in} \dom(h_1)\}
\\
  r_b \defeq \{ (h_1,h_2) \mid 1 {\in} \dom(h_2) \}
\end{array}
$$
are preserved by the modules: the relational meaning of $1 \ipsto\blank * a
\,\vee\, 1 \ipsto\blank * b$ is empty; note that separating conjunctions
binds more tightly than conjunction and disjunction. Hence the client on the
left should give the same result for both modules and, indeed, both
$\init_1;\final_1$ and $\init_2;\final_2$ are the same skip command.

The client on the right also first calls $\init$ and then uses the
rule of consequence. But this time our results say that a true
implication $(1\ipsto\blank \wedge a * b) \implies (1 \ipsto\blank * a
\vee 1 \ipsto\blank * b)$ in the rule of consequence does \emph{not}
lift to an implication between relational meanings of the assertions:
the pair of heaps $([1 \mapsto 0],[1 \mapsto 0])$ belong to the left
hand side but possibly not to the right if the pair $([],[1 \mapsto
0])$ is in the interpretation of $a$ and the pair $([1 \mapsto 0],[])$
is in the interpretation of $b$; see Example \ref{example:fan} for
details. Because of this failure, the proof of the client does not
ensure representation independence. In fact, the client can indeed
distinguish between the two module implementations --- when the client
is executed with the first module implementation, the final heap maps
address $1$ to $1$, but when the client is executed with the second,
the final heap maps address $1$ to $2$.

Note that we phrase the lifting only in terms of
semantically true implications, without referring to how
they are proved. By doing so, we make our results relevant
to automatic tools that use the semantic model of separation logic
to prove implications, such as the ones based on shallow embeddings of
the assertion logic~\cite{BirkedalL:ynot-conf,Dockins-aplas09}.

To sum up, the question of whether representation independence holds
for a client with a proof comes down to whether, in the proof, the
implications used in the rule of consequence can be lifted to a
relational interpretation. In this paper, we give a sound and, in a
certain sense, complete characterization of when that holds.

It is appropriate to remark already here, that although we extend our
assertions with assertion variables we also restrict them to contain
neither ordinary nor separating implication. And, in the end, we
consider only a fragment of those. Details are given in Sections 2 
and 4; here we just remark that the assertions we do study are not
unlike the ones considered in the tools mentioned in the beginning of
this introduction, in particular \jstar. Also we use intuitionistic
separation logic as we envision a language with garbage collection;
this, too, is in line with the \jstar{} tool.

The rest of the paper is organized as follows:
\begin{desCription}
\item\noindent{\hskip-12 pt\bf Sections \ref{sec:domain} and
  \ref{sec:assertion}:}\ give the meanings of
assertions, both the standard and relational meanings. Indeed, we give, for
any $n \in \pinteger$, the $n$-ary meaning of assertions as $n$-ary relations
on the set of heaps. These relations are intuitionistic, i.e., they are upward
closed relations with respect to heap extension.

\item\noindent{\hskip-12 pt\bf Section \ref{section:liftingtheoremsandcompleteness}:}\ contain the main
technical contributions of the paper. We give, for assertions of a particular
form, a sound and, in a certain sense, complete answer to the question of
when we may lift implications between assertions from the standard, unary
meaning to the binary meaning.

\item\noindent{\hskip-12 pt\bf Section \ref{section:higheraritiesandparametricity}:}\ has the curious
spinoff result that an implication between assertions holds for arbitrary
arity if and only if it holds for reasons of parametric polymorphism in a
particular sense.

\item\noindent{\hskip-12 pt\bf Section \ref{sec:parametricity}:}\ returns to the main line of
development; this is where we show that a client-side proof yields
representation independence if it uses the rule of consequence only with
implications that lift.

\item\noindent{\hskip-12 pt\bf Section \ref{section:conclusionanddiscussion}:}\ concludes the paper.
\end{desCription}

Proofs are found in the appendices, the main text only give details
for a few examples.




\section{Semantic Domain}\label{sec:domain}
In the following section we will define the meaning of an assertion to
be an $n$-ary relation on heaps. To formalize this relational meaning,
we need a semantic domain $\irel_n$ of relations, which we define
and explain in this section.

Let $\heap$ be
the set of finite partial functions from positive integers
to integers (i.e., $\heap \,\defeq\, \pinteger \to_\fin \integer$),
ranged over by $f,g,h$. This is a commonly used set for modelling heaps in
separation logic, and it has a partial commutative
monoid structure $([],\cdot)$, where $[]$ is
the empty heap and the $\cdot$ operator combines disjoint heaps:
$$
[] \defeq \emptyset,
\quad\;
f \cdot g
\defeq
\left\{\begin{array}{@{}l@{\;\;}l@{}}
f \cup g & \mbox{if $\dom(f)\,{\cap}\, \dom(g) \,{=}\, \emptyset$}
\\
\mbox{undefined} & \mbox{otherwise}
\end{array}\right.
$$
The operator $\cdot$ induces a partial order
$\sqsubseteq$ on $\heap$, modelling heap extension,
by $f \,{\sqsubseteq}\, g$ iff $g = f \,{\cdot}\, h$ for some $h$.

We also consider the $+$ operator for combining
possibly-overlapping but consistent heaps, and the $-$ operator for subtracting
one heap from another:
$$
\begin{array}{@{}l@{}}
f + g \;\defeq\;
\left\{\begin{array}{@{}l@{\;}l@{}}
f \cup g & \mbox{if $\forall l {\in} \dom(f) \cap \dom(g).\, f(l) {=} g(l)$}
\\
\mbox{undefined} & \mbox{otherwise}
\end{array}\right.
\\[2ex]
(f - g)(l)  \;\defeq\;
\left\{\begin{array}{@{}l@{\;}l@{}}
f(l) & \mbox{if $l \in \dom(f)\setminus \dom(g)$}
\\
\mbox{undefined} & \mbox{otherwise}
\end{array}\right.
\end{array}
$$

We call an $n$-ary relation $r  \subseteq \heap^n$
\emph{upward closed} iff
$(f_1,\ldots,f_n) \in r \wedge
(\forall i.\; f_i \sqsubseteq g_i)
\implies
(g_1,\ldots,g_n) \in r$.
\begin{defi}
$\irel_n$ is the family of upward closed $n$-ary relations on heaps.
\end{defi}
Note that $\irel_1$ consists of upward closed \emph{sets} of heaps,
which are frequently used to interpret assertions in
 separation logic for garbage-collected languages. We call
elements of $\irel_1$ \emph{predicates} and denote them by $p,q$.

For every $n \geq 1$, domain $\irel_n$ has a complete lattice structure:
join and meet are given by union and
intersection, bottom is the empty relation,
and top is $\heap^n$.  The domain also has a semantic separating
conjunction connective defined by
$$
\begin{array}{@{}r@{}c@{}l@{}}
(f_1,..,f_n) \,{\in}\, r \,{*}\, s
& \defiff &
\exists (g_1,..,g_n) \,{\in}\, r.\;
\exists (h_1,..,h_n) \,{\in}\, s.
\\
&&
\,
(g_1,..,g_n) \cdot
(h_1,..,h_n)  =
(f_1,..,f_n).
\end{array}
$$
Here we use the component-wise extension of $\cdot$ for tuples. Intuitively,
a tuple is related by $r * s$ when it can be split into two disjoint tuples,
one related by $r$ and the other by $s$.

The domain $\irel_1$ of predicates is related to $\irel_n$ for every $n$,
by the map~$\diag_n \,\defeq\, \lambda p. \{(f,\ldots,f) \mid f \in p\}^\uparrow$,
where $\uparrow$ is the upward closure on relations. Note
that each predicate is turned into an $n$-ary identity relation on $p$
modulo the upward closure. This map behaves well with respect
to the structures discussed on $\irel_1$ and $\irel_n$, as expressed
by the lemma below:
\begin{lem}
\label{lem:diag-preservation}
Function $\diag_n$ preserves the complete lattice structure
and the $*$ operator.
\end{lem}

For every $n \geq 1$, the domain $\irel_n$ has further structure: it
has standard, semantic separating implication and upwards-closed
implication; as such, it is a complete BI algebra
\cite{Biering:Birkedal:Torp-Smith:07}. Unfortunately, the above lemma
fails for both implications. And that lemma is the pivot of the
upcoming results; it is the basic link between the unary and binary
(and n-ary) readings of assertions. This is why we leave out these
connectives in our assertions in the next section; it is a fundamental
limitation in our approach.

\section{Assertions and Relational Semantics}\label{sec:assertion}
Let $\var$ and $\avar$ be disjoint sets of normal variables
$x,y,...$ and assertion variables $a,b,...$, respectively. Our
assertions $\varphi$ are from separation logic, and
they conform to the following grammar:
\[
\begin{array}{@{}r@{}c@{}l@{}}
E & \;::=\; &
x \,\mid\, 0 \,\mid\, 1 \,\mid\, E+E \,\mid\, \ldots
\qquad
P \;{::=}\;
E \,\ipsto\, E
\,\mid\,
\ldots
\\
\varphi & \;::=\; &
P
\,\mid\,
a
\,\mid\,
\varphi * \varphi
\,\mid\,
\true
\,\mid\,
\varphi \wedge \varphi
\,\mid\,
\false
\,\mid\,
\varphi \vee \varphi
\\
& \,\mid\, &
\forall x.\, \varphi
\,\mid\,
\exists x.\, \varphi
\end{array}
\]
In the grammar, $E$ is a heap-independent expression, and $P$ is a
primitive predicate, which in the standard interpretation denotes an
upward closed set of heaps. For instance, $E\ipsto E'$ means heaps
containing cell $E$ with contents $E'$. The dots in the grammar
indicate possible extensions of cases, such as multiplication for $E$
and inductive predicates for $P$. We will use the abbreviation
$E\ipsto\blank$ for $\exists y.E\ipsto y$.

An assertion $\varphi$ is given a meaning $\semn{\varphi}_{\eta,\rho}
\in \irel_n$ as an $n$-ary relation on heaps, where the arity $n$ is a
parameter of the interpretation. Here environment $\eta$ maps normal
variables in $\varphi$ to integers, and $\rho$ maps assertion
variables in $\varphi$ to $n$-ary relations in $\irel_n$. When
$\varphi$ does not contain any assertion variables, we often omit
$\rho$ and write $\semn{\varphi}_\eta$, because the meaning of
$\varphi$ does not depend on $\rho$. We will make use of unary and
binary semantics most places, but in Section
\ref{section:higheraritiesandparametricity} we will explore higher
arities as well.

We define the semantics of $\varphi$, using the complete lattice
structure and the $*$ operator of the domain $\irel_n$; see
Figure~\ref{fig:semantics-assertion}. Note that the relational
semantics of primitive predicates is defined by embedding their
standard meanings via $\diag_n$. In fact, this embedding relationship
holds for all assertions without assertion variables, because
$\diag_n$ preserves the semantic structures of the domains
(Lemma~\ref{lem:diag-preservation}):
\begin{lem}\label{lem:diag-semantics}
For all $\varphi$ and $\eta,\rho,\rho'$,
if $\diag_n(\rho(a)) = \rho'(a)$ for every $a \in \avar$, we
have that $\diag_n(\semun{\varphi}_{\eta,\rho}) = \semn{\varphi}_{\eta,\rho'}$.
\end{lem}

\begin{figure}[t]
\hrule
$$
\begin{array}{@{}l@{\quad}l@{}}
\semn{P}_{\eta,\rho}
\,\defeq\,
\diag_n(\semprim{P}_{\eta})
&
\semn{\varphi *\psi}_{\eta,\rho}
\,\defeq\,
\semn{\varphi}_{\eta,\rho} * \semn{\psi}_{\eta,\rho}
\\
\semn{a}_{\eta,\rho}
\,\defeq\,\rho(a)
&
\semn{\varphi \wedge \psi}_{\eta,\rho}
\,\defeq\,
\semn{\varphi}_{\eta,\rho} \cap \semn{\psi}_{\eta,\rho}
\\
\semn{\true}_{\eta,\rho} \,\defeq\, \heap^n
&
\semn{\varphi \vee \psi}_{\eta,\rho}
\,\defeq\,
\semn{\varphi}_{\eta,\rho} \cup \semn{\psi}_{\eta,\rho}
\\
\semn{\false}_{\eta,\rho} \,\defeq\, \emptyset
&
\semn{\forall x.\varphi}_{\eta,\rho}
\,\defeq\,
\bigcap_{v \in \integer}\semn{\varphi}_{\eta[x{\mapsto}v],\rho}
\\
&
\semn{\exists x.\varphi}_{\eta,\rho}
\,\defeq\,
\bigcup_{v \in \integer}\semn{\varphi}_{\eta[x{\mapsto}v],\rho}
\end{array}
$$
where $\semprim{P}_\eta$ is the standard semantics of $P$ as
an upward closed set of heaps, which satisfies:
\vspace{-1mm}
$$
\semprim{E\ipsto F}_\eta
\;=\;
\{f \mid \sem{E}_\eta \in \dom(f) \;\wedge\; f(\sem{E}_\eta) = \sem{F}_\eta \}.
$$
\hrule
\caption{Interpretation of Assertions}
\label{fig:semantics-assertion}
\end{figure}

We write $\varphi \models^n \psi$ to mean that
$\semn{\varphi}_{\eta,\rho} \subseteq \semn{\psi}_{\eta,\rho}$
holds for all environments $\eta,\rho$. If $n{=}1$, this reduces
to the standard semantics of assertions in separation logic. We
will use the phrase ``$\varphi {\implies} \psi$ is $n$-ary valid''
to mean that $\varphi \models^n \psi$ holds. In addition, we write
$\varphi \models^n_\eta \psi$ for a fixed $\eta$ to mean that
$\semn{\varphi}_{\eta,\rho} \subseteq \semn{\psi}_{\eta,\rho}$
holds for all environments $\rho$; we say that ``$\varphi \implies
\psi$ is $n$-ary $\eta$-valid'' if this is true.


\section{Lifting Theorems and Completeness}\label{section:liftingtheoremsandcompleteness} 

We call an assertion $\varphi$ \emph{simple} if it is of the form
$(\bigvee_{i{=}1}^I \bigwedge_{j=1}^J \varphi_{(i,j)} *
\vec{a}_{(i,j)})$, where $\vec{a}_{(i,j)}$ is a vector of assertion
variables and $\varphi_{i,j}$ is an assertion not containing any
assertion variables. We will consider the question of lifting an
implication between simple assertions $\varphi,\psi$ to a binary
relational interpretation: when does $\varphi \models^1 \psi$ imply
that $\varphi \models^2\psi$?

The simple assertions are a fragment of the assertions considered in
the above section: simple assertions are not, in general, closed under
separating conjunction as the latter does not distribute over
conjunction, nor are quantified simple assertions necessarily simple.
The divide, however, between simple and non-simple assertions is not
deep. The forthcoming completeness result is intimately connected to
the form of the assertions, but it is very possible that the basic
ideas from lifting could be applied to a larger fragment. We have not
considered that to any extent, however. It is worth mentioning that
all assertions considered in Section 1 are simple. On the other hand,
for assertions $\varphi_1$, $\varphi_2$ and $\varphi_3$ with no
assertion variables and assertion variables $a_1$,$a_2$ and $a_3$, we
do not, in general, have simplicity of an assertion like
\[
  \big[(\varphi_1 * a_1) \wedge (\varphi_2 * a_2)\big] * \big[\varphi_3 * a_3\big].
\]
It should be noted, that simple assertions include most of the
important aspects of the fragments of separation logic used by
automatic program analysis tools. For instance, if we ignore so called
primed variables (which correspond to existentially-quantified
variables), the original SpaceInvader uses separation-logic formulas
of the form $\bigvee_{i{=}1}^I (P_{i,1} * \ldots *
P_{i,k_i})$~\cite{Distefano-TACAS06}, and its most recent extension
for handling a particular class of graph-like data structures uses
$\bigwedge_{j = i}^J \bigvee_{i{=}1}^I (P_{i,j,1} * \ldots *
P_{i,j,k_{i,j}})$~\cite{LeeYangRasmussemCAV11}. Note that in both
cases, either formulas are already simple or they can be easily
transformed to equivalent simple formulas. The assertions used by the
\jstar{} tool \cite{distefano2008jstar} has neither ordinary
implication, separating implication nor ordinary conjunction and only
quite restricted use of quantifiers. Since proofs obtained from such
tools are one target of our results, we argue that the restrictions
imposed on assertions are not unreasonable in terms of usage.

The starting point of our analysis is to realize that it is sufficient to study
implications of the form:
\begin{equation}
\label{eqn:implication-form}
\bigwedge_{i=1}^M \varphi_i * a_{i,1} * \cdots * a_{i,M_i}
\implies
\bigvee_{j=1}^N \psi_j * b_{j,1} * \cdots * b_{j,N_j}
\end{equation}
where $\varphi_i$'s and $\psi_j$'s do not contain assertion
variables, and no assertion variables occur only on the right hand
side of the implication.
\begin{lem}\label{lemma:implication-reduction}
There is an algorithm taking simple assertions $\varphi,\psi$ and
returning finitely many implications $\{\varphi^l {\implies}
\psi^l\}_{l \in L}$, such that (a) $\varphi^l {\implies} \psi^l$ has
the form~\eqref{eqn:implication-form} for all $l \in L$, and (b) for
any $n \in \{1,2\}$, we have that $\varphi \models^n \psi$ holds iff
$\varphi^l \models^n \psi^l$ holds for all $l \in L$.
\end{lem}
The algorithm in the lemma is given in Appendix~\ref{appendix:proof-implication-reduction}.

Thus, in this section, we will focus on lifting implications of
the form \eqref{eqn:implication-form}. Specifically, we will give
a \emph{complete} answer to the following question: Given one such
implication that is $\eta$-valid in the unary interpretation for
some environment $\eta$, can we decide if the implication is
$\eta$-valid in the binary interpretation \emph{merely} by
inspection of the layout of the assertion variables? The answer
will come in two parts. The first part, in
Section~\ref{subsec:layout-lift}, provides three lifting theorems,
each of which has a criterion on the variable layout that, if met,
implies that $\eta$-validity may be lifted regardless of the
$\varphi_i$'s and $\psi_j$'s. The second part, in
Section~\ref{subsec:completeness}, is a completeness theorem; it
states that if the variables fail the criteria of all three
lifting theorems then there are choices of $\varphi_i$'s and
$\psi_j$'s with no variables such that we have unary but not
binary validity.

This approach has pros and cons. Assume that we have an implication of
the aforementioned form that is valid in the unary interpretation, and
we would like to know if it is valid in the binary interpretation too.
Trying out the layout of the variables against the criteria of the
three lifting theorems is an easily decidable and purely syntactical
process -- and if it succeeds then we have binary validity. If it
fails, however, we are at a loss; we know that there are $\varphi_i$'s
and $\psi_j$'s with the same variable layout such that lifting fails
but we do not learn anything about our concrete implication. There is,
however, an alternate use of the theory below if the lifting criteria
fail; we will elaborate on that in Section
\ref{sec:parametricity}.


\subsection{Notation} \label{subsection:notation}
We need some notation that will accompany us throughout this
section. Consider an implication of the form
\eqref{eqn:implication-form}. Let $V = \bigcup_{i=1}^M \{a_{i,1},
\ldots,a_{i,M_i}\}$ be the set of all left hand side assertion
variables, these include the right hand side assertion variables
too by assumption. Define $\Pi:\{1, \ldots, M\} \to \nat^V$ and
$\Omega: \{1, \ldots, N\} \to \nat^V$ by the following:
$$
\Pi(i)(c)\,\defeq\, |\{k \mid a_{i,k} \equiv c\}|,
\quad
\Omega(j)(c) \,\defeq\, |\{k \mid b_{j,k} \equiv c\}|.
$$
These functions give vectors of assertion variable counts for each
conjunct and disjunct. For $1 \leq i \leq M$ and $1 \leq j \leq N$
we write $\Pi(i) \geq \Omega(j)$ if we have $\Pi(i)(c) \geq
\Omega(j)(c)$ for each variable $c \in V$, i.e., if conjunct $i$
has the same or a greater number of occurrences of all variables
than disjunct $j$. We write $\Pi(i) \ngeq \Omega(j)$ if this
fails, i.e., if there is $c \in V$ such that $\Pi(i)(c) <
\Omega(j)(c)$. If a conjunct, say conjunct $i$, has no variables,
i.e., if $\Pi(i)(c) = 0$ holds for all $c \in V$, then we say it
is \emph{empty}; the same goes for the disjuncts.

We shall write $-$ to denote $\exists n,m.\ n \ipsto
m$, meaning heaps with at least one cell. On the semantic side, we
write $[m]$ for $m \in \pinteger$ to denote the heap that stores 0
at location $m$ and nothing else. For $m_0, ..., m_n \in
\pinteger$ different we write $[m_0, ..., m_n]$ for
$[m_0]\cdot ... \cdot [m_n]$.

Finally we introduce a piece of sanity-preserving graphical notation. We
depict an implication of the form \eqref{eqn:implication-form} as a complete
bipartite graph with the conjuncts lined up on the left hand side and the
disjuncts on the right hand side. For any $1 \leq i \leq M$ and any $1 \leq j
\leq N$ we draw a solid line from conjunct $i$ to disjunct $j$ if $\Pi(i)
\geq \Omega(j)$. We label that line with all the $c \in V$ such that
$\Pi(i)(c) > \Omega(j)(c)$ if indeed there are any such. If, on the other
hand, $\Pi(i) \ngeq \Omega(j)$ then we draw a dashed line instead and label
it with all the $c \in V$ such that $\Pi(i)(c) < \Omega(j)(c)$. Note that the
drawing of edges depend solely on the layout of the variables; the
$\varphi_i$'s and $\psi_j$'s have no say in the matter. As an example, the implication
\[
  1 \ipsto \blank \wedge a * b \implies 1 \ipsto \blank * a \vee 1 \ipsto \blank * b,
\]
which we shall look into in Example \ref{example:fan}, is depicted as follows:
 \[\xymatrix@R=7ex@C=1ex{
   1\ipsto \blank & \bullet \ar@{--}[rrr]_a \ar@{--}[drrr]_<<b &&& \bullet & 1 \ipsto \blank * a \\
   a \ast b &  \bullet \ar@{-}[rrr]_a \ar@{-}[urrr]^<<b &&& \bullet & 1\ipsto\blank * b}
\]
With a little experience, it is quite easy to check the conditions of
the upcoming lifting theorems by looking at the corresponding graph;
the graphs expose the structure of the assertions important to the
proofs.

\subsection{Strategy}

We give a brief strategic overview before the onslaught. Consider an
implication of the form \eqref{eqn:implication-form}. If the layout of
the variables satisfy (at least) one of three upcoming criteria then
we know this: unary $\eta$-validity holds only if it holds for
`obvious reasons'. The latter is captured precisely in the
Parametricity Condition but, loosely, it says that there are $1 \leq i
\leq M$ and $1 \leq j \leq N$ such that $\phi_i \implies \psi_j$ is
$\eta$-valid in the unary interpretation and such that $\Pi(i) \geq
\Omega(j)$. This is sufficiently parametric in the treatment of
assertion variables that it immediately implies binary $\eta$-validity
and even $n$-ary $\eta $-validity for any $n$.

The three criteria, as given in the next subsection, are rather
technical; each is what it takes for proof idea of the corresponding
lifting theorem to go through. They are complete, however: if the
implication fails all three criteria then there are choices of
$\varphi_i$'s and $\psi_j$'s such that unary $\eta$-validity holds for
`non-obvious reasons'; in particular such that binary $\eta$-validity
fails. Non-obvious reasons comes down to exploiting the limited arity
in different ways; we elaborate on that in Subsection
\ref{subsec:completeness}.

\subsection{Layouts that Lift}\label{subsec:layout-lift}
The following is a first example of a layout of variables that
ensure that for any choice of $\varphi_i$'s and $\psi_j$'s we get
that unary $\eta$-validity of the implication yields binary
$\eta$-validity. That it holds is a consequence of Theorem
\ref{theorem:shadow} but we have spelled out a concrete proof that
will serve as a guide to the further development.

\begin{exa}[Shadow-Lift] \label{example:shadow}
{\rm For any four assertions $\varphi_1, \varphi_2, \psi_1,
\psi_2$ with no assertion variables and any appropriate
environment $\eta$ we have that unary $\eta$-validity of the
following implication implies binary $\eta$-validity:
 \[\xymatrix@R=7ex@C=1ex{
   \varphi_1 * a * b& \bullet \ar@{--}[rrr]^b \ar@{--}[drrr]_<<b &&& \bullet & \psi_1 * a  * b*b\\
   \varphi_2 * a * b*b &  \bullet \ar@{-}[rrr]_a \ar@{-}[urrr] &&& \bullet & \psi_2 *b*b
 }\]
Assume that we have unary $\eta$-validity. Before we go on to
consider the binary case we derive a simple unary consequence that
does not involve assertion variables: For any $h \in \heap$ with
subheaps $h_1 \sqsubseteq h$ and $h_2 \sqsubseteq h$ such that
$h_1 \in \semun{\varphi_1}_\eta$ and $h_2 \in
\semun{\varphi_2}_\eta$ we get that $h_2 \in \semun{\psi_1}_\eta$
or that $h_2 \in \semun{\psi_2}_\eta$.

To prove this, let $h$, $h_1$ and $h_2$ be as assumed. We build
$\rho:\{a,b\} \to \irel_1$ by letting $\rho(a) = \heap$ and
letting $\rho(b)$ be the following union of sets of heaps:
  \[\{(h-h_1) \cdot [n, n+1]\}^\uparrow \cup \{(h-h_2)\cdot[n]\}^\uparrow \cup \{[n+1]\}^\uparrow\]
where $n = \max(\dom(h) \cup \{0\})+1$. It is now immediate that
$h \cdot[n,n+1]$ lies in the interpretation of both conjuncts since
\[
h_1 \cdot (h-h_1)\cdot[n,n+1] = h \cdot[n,n+1] = h_2 \cdot (h-h_2)\cdot[n]\cdot[n+1],
\]
and so by our assumption on the original implication it must lie in
the interpretation on one of the disjuncts too. Suppose that we have
\[
h \cdot [n,n+1] \in \semun{\psi_1*a*b*b}_{\eta,\rho} =
\semun{\psi_1}_\eta*\rho(b)*\rho(b),
\]
where the equality holds because $\rho(a) = \heap$ is the unit for $*$.
We then write $h \cdot [n,n+1] = g_1 \cdot g_2 \cdot g_3$ for $g_1
\in \semun{\psi}_\eta$ and $g_2, g_3 \in \rho(b)$. But as $g_2$ and $g_3$
have disjoint domains we must have $(h-h_2)\cdot[n] \sqsubseteq g_2$
and $[n+1] \sqsubseteq g_3$ or the version with $g_2$ and $g_3$
swapped. In any case we have that
$$
\begin{array}{@{}r@{}c@{}l@{}}
\dom(g_1)
&
\;=\;
&
\dom(h \cdot[n, n{+}1]) \setminus (\dom(g_2 \cdot g_3))
\\
&
\;\subseteq\;
&
\dom(h \cdot[n, n{+}1]) \setminus (\dom(h{-}h_2)\,{\cup}\, \{n, n{+}1\})
\\
&
\;=\;
&
\dom(h_2).
\end{array}
$$
But then we have $g_1
\sqsubseteq h_2$ and since $g_1 \in \semun{\psi_1}_\eta$ we get $h_2 \in
\semun{\psi_1}_\eta$ too. If we have $h \cdot [n,n+1] \in
\semun{\psi_2*b*b}_{\eta,\rho}$ we proceed similarly.

The above short proof is the crux of the example. It implies unary
$\eta$-validity -- this we knew already -- but also the binary
$\eta$-validity. To see this, we pick an arbitrary environment
$\rho:\{a,b\} \to \irel_2$, we take arbitrary $(h_1, h_2) \in
\sembin{\varphi_1*a*b \wedge \varphi_2*a*b*b}_{\eta,\rho}$ and we aim
to prove that $(h_1, h_2) \in \sembin{\psi_1*a*b*b \vee
\psi_2*b*b}_{\eta,\rho}$ too. We split $(h_1, h_2)$ according to the
conjuncts. Because of Lemma~\ref{lem:diag-semantics} and the upward
closedness condition of $\irel_2$, we can write \[(h_1, h_2) = (g^1,
g^1) \cdot (g_1^2, g_2^2) \cdot (g_1^3, g_2^3)\] for $g^1 \in
\semun{\varphi_1}_\eta$, $(g_1^2,g_2^2) \in \rho(a)$ and $(g_1^3,
g_2^3) \in \rho(b)$. Also we can write \[(h_1, h_2) = (f^1, f^1) \cdot
(f_1^2, f_2^2) \cdot (f_1^3, f_2^3) \cdot (f_1^4, f_2^4)\] for $f^1
\in \semun{\varphi_2}_\eta$, $(f_1^2, f_2^2) \in \rho(a)$ and $(f_1^3,
f_2^3), (f_1^4, f_2^4) \in \rho(b)$. But now $g^1+f^1$ with subheaps
$g^1$ and $f^1$ meet the conditions of the unary consequence from
above, and so we get $f^1 \in \semun{\psi_1}_\eta$ or $f^1 \in
\semun{\psi_2}_\eta$ and the second splitting of $(h_1, h_2)$ shows
that $(h_1, h_2)$ lie in the binary interpretation of the first or
second disjunct, respectively. Notice that neither $g^1 \in
\semun{\psi_1}_\eta$ nor $g^1 \in \semun{\psi_2}_\eta$ would have
worked since the first conjunct has too few variables, i.e., $\Pi(1)
\ngeq \Omega(1)$ and $\Pi(1) \ngeq \Omega(2)$ \qed }
\end{exa}

The simple idea justifies the odd choice of name: we attach to each
occurrence of $b$ in the conjuncts a `shadow' in such a way that any
two shadows from different conjuncts overlap. This means that the two
occurrences of $b$ in, say, the first disjunction must correspond to
occurrences of $b$ in the same conjunct; in particular that conjunct
must have at least two occurrences. We attach no shadow to $a$,
instead we remove $a$ by instantiating to $\heap$; this is because
the second disjunct lacks an occurrence of $a$ and hence we may fail
to `peel off' the entire shadow. Had $a$ occurred as the single label
of a dashed line, this removal would have `introduced' a solid line
and the approach would fail.

Generalizing the unary consequence that served as the crucial
stepping stone in the above example we arrive at the following
condition on our implications:

\begin{defi}[Parametricity Condition]
\label{def:parametricitycondition} Assume that we have an implication
of the form \eqref{eqn:implication-form} and an appropriate
environment $\eta$. We say that the Parametricity Condition is
satisfied if, for all $h, h_1, \ldots, h_M \in \heap$ with $h_i
\sqsubseteq h$ and $h_i \in \semun{\varphi_i}_\eta$ for all $1 \leq i
\leq M$, it is is the case that one (or both) of the following
conditions hold:
\begin{enumerate}[(1)]
  \item There are $1 \leq i \leq M$ and $1\leq j \leq N$
  such that $h_i \in \semun{\psi_j}_\eta$ and $\Pi(i) \geq \Omega(j)$.
  \item There is $1 \leq j \leq N$ such that $h \in \semun{\psi_j}_\eta$ and the $j$-th disjunct is empty.
\end{enumerate}
\end{defi}
Note that specializing the Parametricity Condition, henceforth
just the PC, to an implication of the form treated in the above
example yields the stated unary consequence because no disjuncts
are empty. The second option in the PC will be motivated later.

We emphasize that the PC may hold or may fail for any given
combination of an implication and environment $\eta$. But if it
holds then we have binary $\eta$-validity; the proof in case of
the first option of the PC is an easy generalization of the latter
half of the above example:

\begin{prop} \label{proposition:gycimpliesbinary}
The PC implies binary $\eta$-validity.
\end{prop}

We arrive now at the first lifting theorem. It is a generalization
of the former half of Example \ref{example:shadow}; the proof of
the theorem has a lot more details to it than the example but the
overall idea is the same. The theorem states a criterion on the
layout of the variables that, if met, means that unary
$\eta$-validity implies the PC and hence also binary
$\eta$-validity. The criterion is, loosely, that we can remove all
variables that occur as labels of solid lines without introducing
new solid lines and without emptying any disjuncts:

\begin{thm}[Shadow-Lift] \label{theorem:shadow}
Unary $\eta$-validity of an implication implies the PC if each dashed
line is labeled with at least one variable which is not a label on a
solid line and each disjunct has an occurrence of a variable that is
not a label on a solid line. Spelling it out in symbols, we require,
with $L = \{(i,j) \mid 1\leq i \leq M \wedge 1 \leq j \leq N\}$, that
\[
\begin{array}{@{}l@{}}
 \forall (i,j) \in L.\ \; \Pi(i) \ngeq \Omega(j) \implies {}
\\
  \;\;\; \exists c \in V.\ \; \Pi(i)(c) < \Omega(j)(c) \,\wedge\, {}
\\
  \;\;\;\;\; \bigl(\forall (k,l) \in L.\ \; \Pi(k) \geq \Omega(l) \implies  \Pi(k)(c) = \Omega(l)(c)\bigr)
\end{array}
\]
and
$$
\begin{array}{@{}l@{}}
   \forall 1 \leq j \leq N.\ \; \exists c \in V.\ \;
   \Omega(j)(c) > 0\,\wedge\, {}
\\
   \;\;\;
   \bigl(\forall (k,l) \in L.\ \; \Pi(k) \geq \Omega(l) \implies
   \Pi(k)(c) = \Omega(l)(c)\bigr).
\end{array}
$$
\end{thm}

As motivation for the next lifting theorem, we note that the
variable layout criterion of the above theorem fails if one or
more disjuncts are empty. Correspondingly, we never touch upon the
second option of the PC. But there are variable layouts with
empty disjuncts that ensure lifting:

\begin{exa}[Balloon-Lift] \label{example:balloon}
{\rm
For any four assertions $\varphi_1, \varphi_2, \psi_1,
\psi_2$ with no assertion variables and any appropriate
environment $\eta$ we have that unary $\eta$-validity of the
following implication implies binary $\eta$-validity:
 
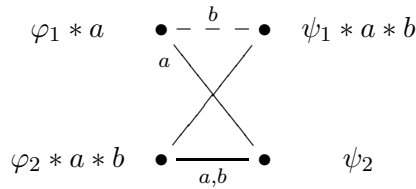
\[\xymatrix@R=7ex@C=1ex{
   \varphi_1 * a & \bullet \ar@{--}[rrr]^b \ar@{-}[drrr]_<<a &&& \bullet & \psi_1 * a * b \\
   \varphi_2 * a * b &  \bullet \ar@{-}[rrr]_{a,b} \ar@{-}[urrr] &&& \bullet & \psi_2
 }\]
Assume unary $\eta$-validity. As in Example \ref{example:shadow}
we derive a unary consequence as an intermediate result: For any $h
\in \heap$ with subheaps $h_1 \sqsubseteq h$ and $h_2 \sqsubseteq
h$ such that $h_1 \in \semun{\varphi_1}_\eta$ and $h_2 \in
\semun{\varphi_2}_\eta$ we have that either $h_2 \in
\semun{\psi_1}_\eta$ or $h \in \semun{\psi_2}_\eta$ .

To prove this, let $h$, $h_1$ and $h_2$ be as assumed. We
construct $\rho : \{a,b\} \to \irel_1$ by letting
$\rho(a) = \heap$ and $\rho(b) = \{h - h_2\}^\uparrow$. We get that
$$
\begin{array}{@{}l@{}}
h \,\sqsupseteq\, h_1 \,\in\, \semun{\varphi_1}_\eta \,=\, \semun{\varphi_1}_\eta * \heap \,=\, \semun{\varphi_1 * a}_{\eta,\rho},
\\[1ex]
\mbox{and}
\\[1ex]
h
\begin{array}[t]{@{}l@{}}
  {} = h_2 \cdot (h - h_2)
\\
  {} \in \semun{\varphi_2}_\eta * \rho(b) = \semun{\varphi_2}_\eta * \heap * \rho(b) = \semun{\varphi_2  *a*b}_{\eta,\rho}.
\end{array}
\end{array}
$$
This means that $h$ must lie in the interpretation of one of the
disjuncts. If it is the first, we inspect the interpretation and
get that
$$
h = g_1 \cdot g_2 \cdot g_3
$$
for $g_1 \,{\in}\, \semun{\psi_1}_\eta$, $g_2 \,{\in}\, \heap$ and $g_3 \,{\sqsupseteq}\, h {-} h_2$. It means that
$$
\begin{array}{@{}r@{}c@{}l@{}}
   \dom(g_1)
   & {} \,=\, {} & \dom(h) \setminus \dom(g_2 \cdot g_3)
   \,\subseteq\, \dom(h) \setminus \dom(g_3)
\\
   & {} \,\subseteq\, {} & \dom(h) \setminus \dom(h - h_2) \,=\, \dom(h_2)
\end{array}
$$
which implies that $g_1 \sqsubseteq h_2$ and so $h_2 \in
\semun{\psi_1}_\eta$. If, on the other hand, $h$ lies in the
interpretation of the second disjunct then we are done
immediately.

Now we prove the claim of binary $\eta$-validity. We pick an arbitrary
environment $\rho:\{a,b\} \to \irel_2$, we take arbitrary $(h_1, h_2)
\in \sembin{\varphi_1*a \wedge \varphi_2*a*b}_{\eta,\rho}$ and we must
prove that $(h_1, h_2) \in \sembin{\psi_1*a*b \vee
\psi_2}_{\eta,\rho}$ too. We write \[(h_1, h_2) = (g^1, g^1) \cdot
(g_1^2,g_2^2) \] for $g^1 \in \semun{\varphi_1}_\eta$ and
$(g_1^2,g_2^2) \in \rho(a)$, and \[(h_1, h_2) = (f^1, f^1) \cdot
(f_1^2,f_2^2) \cdot (f_1^3,f_2^3) \] for $f^1 \in
\semun{\varphi_2}_\eta$, $(f_1^2,f_2^2) \in \rho(a)$ and
$(f_1^3,f_2^3) \in \rho(b)$. But now $g^1 + f^1$ with subheaps $g^1$
and $f^1$ satisfies the above properties and so we get $f^1 \in
\semun{\psi_1}_\eta$ or $g^1+f^1 \in \semun{\psi_2}_\eta$. If $f^1 \in
\semun{\psi_1}_\eta$ holds then the second splitting of $(h_1,h_2)$
shows that $(h_1, h_2)$ is in the interpretation of the first
disjunct. If $g^1+f^1 \in \semun{\psi_2}_\eta$, we are done too, since
we may write $(h_1, h_2) = (g^1 + f^1, g^1 + f^1) \cdot (e_1, e_2)$
for some $(e_1, e_2) \in \heap^2$ and so $(h_1, h_2)$ lies in the
interpretation $\sembin{\psi_2}_\eta = \Delta(\semun{\psi_2}_\eta)$ of
the second conjunct. \qed }
\end{exa}

Once again, the underlying idea is simple: we attach `shadows' to
occurrences of variables, but this time we stay within the the
original heap. This is quite inhibitory as we may have to use the
empty heap as shadow. Again we remove a variable (in general a set of
variables) by instantiating to $\heap$ but the remaining variable (in
general the remaining set of variables) must satisfy quite restrictive
demands.

Just as we did for Example \ref{example:shadow} we may generalize the
former half of this example yielding Theorem \ref{theorem:balloon}
below. The latter half of the example, on the other hand, constitutes
an example of the approach of the proof of Proposition
\ref{proposition:gycimpliesbinary} in case we run into the second
option of the PC. Note also that specializing the PC to an implication
of the form considered in the example yields the stated unary
consequence.

\begin{thm}[Balloon-Lift] \label{theorem:balloon}
Unary $\eta$-validity of an implication implies the PC if there is
a subset $B \subseteq V$ with the following three properties.
First, each conjunct has at most one occurrence of a variable from
$B$, i.e.,
$$
\forall 1\leq i\leq M.\ \sum_{c \in B} \Pi(i)(c) \leq 1.
$$
Second, each disjunct is empty or has exactly one occurrence of a
variable from $B$, i.e.,
$$
\forall 1\leq j\leq N.\ \sum_{c \in V} \Omega(j)(c) = 0 \,\vee\, \sum_{c\in B} \Omega(j)(c) = 1.
$$
Third,  each dashed line must have a label from $B$. That is,
when $L = \{(i,j) \mid 1\leq i \leq M \wedge 1\leq j \leq N\}$,
$$
 \forall (i,j) \,{\in}\, L.
 \;
 \Pi(i) \ngeq \Omega(j) \implies \exists c \,{\in}\, B.\ \Pi(i)(c) < \Omega(j)(c).
$$
\end{thm}
One thing to note about the theorem is that if we have no empty
disjuncts, none of the variables in the subset $B \subseteq V$ can be
labels of a solid line. In particular, the conditions of Theorem
\ref{theorem:shadow} are met, so the above theorem is really only
useful if one or more disjuncts are empty. A simple but pleasing
observation is that this theorem is applicable if all conjuncts and
all disjuncts have at most a single occurrence of any assertion
variable; in that case, we can just choose $B=V$ above.

The final lifting theorem captures the oddities of the special
case of just one conjunct:
\begin{thm}[Lonely-Lift] \label{theorem:lonely}
Unary $\eta$-validity of an implication implies the PC if there is
just one conjunct, i.e., $M{=}1$, and all lines are solid, i.e.,
$\Pi(1) \,{\geq}\, \Omega(j)$ for all $1\,{\leq}\,j\,{\leq}\,N$.
\end{thm}

\subsection{Completeness}\label{subsec:completeness}
It is now time for examples of implications that do not lift,
i.e., that are valid in the unary interpretation but not in the
binary. The first is based on the following observation: If $h \in
\semun{1 \ipsto\blank }_\eta$ and $h \in p * q$ for $h \in \heap$
and $p, q \in \irel_1$ then we have $h \in \semun{1 \ipsto \blank}_\eta * p$ or $h \in \semun{1 \ipsto \blank}_\eta * q$. This is because
we must have $[1\mapsto n] \sqsubseteq h$ for some $n \in
\integer$ and so writing $h = h_1 \cdot h_2$ with $h_1 \in p$ and
$h_2 \in q$ gives us $[1\mapsto n] \sqsubseteq h_1$ or $[1\mapsto
n] \sqsubseteq h_2$. But this line of argument breaks down if we
change to binary reading. We have, e.g., $([1],[1]) \in \sembin{1
\ipsto\blank}_\eta$ and $([1],[1]) \in \{([1],[])\}^\uparrow *
\{([],[1])\}^\uparrow$ but both $\sembin{1 \ipsto \blank}_\eta *
\{([1],[])\}^\uparrow$ and $\sembin{1 \ipsto \blank}_\eta
* \{([],[1])\}^\uparrow$ are empty. We can recast this as an
implication that cannot be lifted:

\begin{exa}[Fan-Counter] \label{example:fan}
{\rm This implication is valid on the unary but not on the binary
level:
 
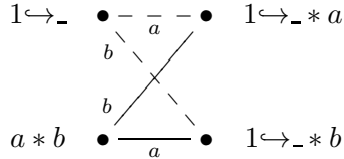
\[\xymatrix@R=7ex@C=1ex{
   1\ipsto \blank & \bullet \ar@{--}[rrr]_a \ar@{--}[drrr]_<<b &&& \bullet & 1 \ipsto \blank * a \\
   a \ast b &  \bullet \ar@{-}[rrr]_a \ar@{-}[urrr]^<<b &&& \bullet & 1\ipsto\blank * b
 }\]
First we argue that the implication holds on the unary level. Let
$\rho:\{a,b\} \to \irel_1$ be an arbitrary environment of upwards
closed sets of heaps to $a$ and $b$. Let $h \in \heap$
be arbitrary and assume that
$$
  h
  \,\in\,
  \semun{1\ipsto\blank \wedge (a \ast b)}_{\eta,\rho}
  \,=\,
  \semun{1\ipsto\blank}_\rho \cap (\rho(a)*\rho(b)).
$$
By the above observation we get either $h \in
\semun{1\ipsto\blank}_\eta * \rho(a)$ or $h \in
\semun{1\ipsto\blank}_\eta * \rho(b)$ which matches the right hand
side of the implication.

Now we move on to prove that the implication fails on the binary
level. Define an environment $\rho:\{a,b\} \to \irel_2$ by
$\rho(a) = \{([1],[])\}^\uparrow$ and $\rho(b) =
\{([],[1])\}^\uparrow$. Then, $\sembin{1\ipsto\blank \wedge a \ast
b}_{\eta,\rho}
  =
 \sembin{1\ipsto\blank}_{\eta,\rho}  \cap (\rho(a) \ast \rho(b))$,
which contains the pair $([1],[1])$. But, as observed, both
disjuncts have empty binary interpretations. \qed }
\end{exa}

An observation of similar nature is that for $p \in \irel_1$ we have
either $p = \heap$ or $p \subseteq \semun{-}_\eta = \{[m {\mapsto} n]
\mid m \in \pinteger, n \in \integer\}^\uparrow$ because if $p \neq
\heap$ then it cannot contain the empty heap. On the binary level,
however, we have $\heap^2 \neq \{([1],[])\}^\uparrow \nsubseteq
\sembin{-}_\eta = \{([m {\mapsto} n],[m {\mapsto} n]) \mid m \in
\pinteger, n \in \integer\}^\uparrow$. One consequence is this:

\begin{exa}[Bridge-Counter] \label{example:bridge}
{\rm This implication is valid on the unary but not on the binary
level:
 
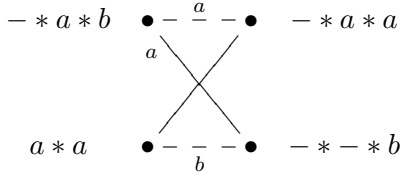
\[\xymatrix@R=7ex@C=1ex{
   - * a * b & \bullet \ar@{--}[rrr]^a \ar@{-}[drrr]_<<a   &&& \bullet & - * a * a \\
   a * a&  \bullet \ar@{--}[rrr]_b \ar@{-}[urrr] &&& \bullet & - * - * b
 }\]
First we argue that the implication holds on the unary level. Let
$\rho:\{a,b\} \to \irel_1$ be an arbitrary environment that
assigns upwards closed sets of heaps to each of the two variables.
We branch on the value of $\rho(a)$.  If $\rho(a) \neq \heap$ then
we have $\rho(a) \subseteq  \semun{-}_\eta$ which again means that
the first conjunct directly implies the second disjunct. If
$\rho(a) = \heap$ holds, we get that
$$
\begin{array}{@{}l@{}l@{}}
  \semun{- \ast a \ast b}_{\eta,\rho}
  \,=\,
  \semun{-}_{\eta,\rho} \ast \heap \ast \rho(b)
  \,=\,
  \semun{-}_{\eta,\rho} \ast  \rho(b)
\\[0.5ex]
  \qquad {} \subseteq\,
  \semun{-}_{\eta,\rho}
  \,=\,
  \semun{-}_{\eta,\rho} * \heap * \heap
  \,=\,
  \semun{- * a * a}_{\eta,\rho}
\end{array}
$$
because $\heap$ is the unit for $\ast$. Hence we get that the
first conjunct implies the first disjunct and we have proved that
the implication holds unarily.

Now we prove that the implication fails on the binary level.
Define an environment $\rho:\{a,b\} \to \irel_2$ by $\rho(a) =
\{([1],[])\}^\uparrow \cup \{([2],[2])\}^\uparrow$ and $\rho(b) =
\heap^2$.  Observe now that $([1,2],[2]) = ([2],[2]) \cdot
([1],[]) \cdot ([],[])$, which implies that $([1,2],[2]) \in
\sembin{- * a * b}_{\eta,\rho}$. From the rewriting $([1,2],[2]) =
([1],[]) \cdot ([2],[2])$, we get $([1,2],[2]) \in \sembin{a *
a}_{\eta,\rho}$ too and so this pair of heaps lies in the
interpretation of the left hand side. But it does not belong to
the interpretation of either disjunct. An easy -- if somewhat
indirect -- way of realizing this is to note that any pair of
heaps in either $\sembin{-}_{\eta,\rho}$ or in $\sembin{a*
a}_{\eta,\rho}$ must have a second component with nonempty domain.
But then any pair of heaps in the interpretation of either
disjunct must have a second component with a domain of at least
two elements. In particular, neither can contain the pair
$([1,2],[2])$. \qed}
\end{exa}

In principle, the above two observations are all that we need to
prove completeness. Or, phrased differently, assume that we have a
layout of variables that fail the criteria of all three lifting
theorems; by applying one of the two observations, we can then
build a concrete implication with that variable layout and with
unary but not binary validity.

Having said that, the territory to cover is huge; the full
completeness proof
is a lengthy and rather technical journey, the details of which do
not provide much insight. We supply it as a series of lemmas in
Appendix \ref{appendix:completeness}; these include
generalizations of Example \ref{example:fan} and Example
\ref{example:bridge} above. If one verifies the lemmas in the
order listed and apply them as sketched then it is feasible, if
not exactly easy, to prove the following:

\begin{thm}[Completeness]  \label{theorem:complete}
If a variable layout meets none of the criteria in Theorems
\ref{theorem:shadow}, \ref{theorem:balloon} and
\ref{theorem:lonely}, then there are choices of $\varphi_i$'s and
$\psi_j$'s with no variables such we have unary but not binary
validity.
\end{thm}

\subsection{Future Work: Supported Assertions} 

By now, we have given a complete division of the possible layouts of
variables into those that lift and those that do not. The divide is
technical; without some understanding of the underlying proofs, it
is hard to get an intuitive feel for it.

One way to simplify would be to consider \emph{supported} assertions.
A $n$-ary relation $r \in \irel_n$ is supported if, for every
$\mathbf{f} \in \heap^n$, it holds that $\mathbf{g_1} \sqsubseteq
\mathbf{f}$ and $\mathbf{g_2} \sqsubseteq \mathbf{f}$ and
$\mathbf{g_1}, \mathbf{g_2} \in r$ together implies the existence of
$\mathbf{g} \in r$ with $\mathbf{g} \sqsubseteq \mathbf{g_1}$ and
$\mathbf{g} \sqsubseteq \mathbf{g_2}$. In an intuitionistic setting,
the supported assertions play the role that the precise assertions do
in a classical, non-intuitionistic setting: they validate reasoning
about the resource invariant in the Hypothetical Frame Rule
\cite{DBLP:journals/toplas/OHearnYR09} and about shared resources in
concurrent separation logic \cite{DBLP:conf/concur/Brookes04}. The
problems we face here appear reminiscent and a natural question is
this: how about restricting assertion variables to supported
assertions?

We have not investigated this in any detail, but initial findings
suggest that this would simplify matters, maybe extensively so. The
counter-example given in Example \ref{example:fan} still holds, so it
is not the case that everything lifts, but the counter-example of
Example \ref{example:bridge} breaks. The central proof of Theorem
\ref{theorem:shadow} uses non-supported assertions; on the other hand,
if $r \in \irel_n$ is supported then either $r * r$ is empty or we
have $r = \heap^n$ so maybe we could restrict to conjuncts with at
most one occurrence of each assertion variable.

Along the same lines, it would be interesting to revisit the
challenges in a classical, non-intuitionistic setting.
This, too, is left for future work with the one comment that the
counter-example given in Example \ref{example:fan} persists.

\section{Higher Arities and Parametricity}\label{section:higheraritiesandparametricity}
We saw in Proposition \ref{proposition:gycimpliesbinary} that the
PC implies binary $\eta$-validity of an implication. It is easy to
show that the PC also implies unary $\eta$-validity, either
directly or by observing that binary implies unary. A natural
question to ask is whether we can reverse this. Example
\ref{example:fan} shows that unary validity does not entail the
PC, because the latter fails for that concrete implication. But as
binary validity fails too, we could hope that binary validity
would enforce the PC. Unfortunately, this is not the case, as
demonstrated by the implication
$$
  1\ipsto\blank \wedge a * a * b
  \implies
  1\ipsto\blank * a \vee 1\ipsto \blank * b.
$$
Here the PC is the same as for Example \ref{example:fan} and hence
still is not true, but we do have binary validity. We do not,
however, have ternary validity but the example could easily be
scaled: having $n$ occurrences of $a$ in the second conjunct means
$n$-ary but not $n+1$-ary validity for any $n \geq 1$. In summary,
we have seen that for any $n \geq 1$ we can have $n$-ary validity
whilst the PC fails.

What does hold, however, is the following:
\begin{thm} \label{theorem:nbigimpliesgyc}
For an implication of the form \eqref{eqn:implication-form}  and an
appropriate environment $\eta$ we have that $n$-ary $\eta$- validity implies
the PC if $n \geq \max\{2,M_1, \ldots, M_M\}$.
\end{thm}
Notice how this fits nicely with the above example: with $n$
occurrences of $a$ we have $n$-ary validity but we need $(n{+}1)$-ary
validity to prove the PC since there is also a single $b$.
The proof is in Appendix
\ref{appendix:higheraritiesandparametricity}, and reuses techniques
from the proofs of Theorems \ref{theorem:shadow} and
\ref{theorem:balloon}.

By an easy generalization of Proposition
\ref{proposition:gycimpliesbinary} we have the following corollary
to the above theorem:
\begin{cor}
The PC holds iff we have $n$-ary $\eta$-validity for all $n \geq
1$.
\end{cor}

This corollary can be read, loosely, as a coincidence between
parametric polymorphism as introduced by Strachey
\cite{Strachey:67} and relational parametricity as proposed by
Reynolds \cite{Reynolds:83}: The PC corresponds to Strachey
parametricity in the loose sense that if it holds, then there is
an approach, parametric in the assertion variables, that produce
right hand side proofs of heap membership from the left hand side
ones: Take a heap, split it along the conjuncts, apply the PC to
the parts in the interpretations of the $\varphi$'s and you are
done, possibly after discarding some variables. This involves no
branching or other intrinsic operations on the assertion
variables, which we are free to discard by our intuitionistic
setup. If, on the other hand, the implication is $\eta$-valid for
arbitrary arity, then it is fair to call it relationally
parametric. Note also that the Examples \ref{example:fan} and
\ref{example:bridge} branch on assertion variable values.

This result is analogous to the conjecture of coincidence
between Strachey parametricity and $n$-ary relational parametricity
for traditional type-based parametricity~\cite[Page~2]{Plotkin:Abadi:93}.

Finally we note that as a consequence of the above corollary we have that the
lifting theorems in the previous section really show that unary validity can
be lifted to validity of arbitrary arity. In some sense, they are stronger
than required for representation independence, for which binary validity
suffices. The authors are unaware of any practical applications of this fact.


\section{Representation Independence} \label{sec:parametricity} 

In this section, we relate our lifting theorems to representation
independence. We consider separation logic with assertion variables
where the rule of consequence is restricted according to our lifting
theorems, and we define a relational semantics of the logic, which
gives a representation independence theorem: all proved clients cannot
distinguish between appropriately related module implementations.

To keep the presentation simple, we omit while-loops and allocation
from the language. Adding the former together with the standard proof
rule is straightforward. Allocation, however, is non-trivial: the
notion of having \emph{one} client using \emph{two} modules will be
hard-coded into our relational reading of the logic, and allocation on
part of the client must give the same address when run with either
module. This fails with standard, non-deterministic allocation; it was
resolved earlier, however, by Birkedal and Yang
\cite{Birkedal:Yang:07} using a combination of FM sets and
continuations and that approach is applicable here too.

We consider commands $C$ given by the grammar:
$$
C
\,::=\,
k \mid
[E]{:=}E \mid
\klet\,y{=}[E]\,\kin\,C \mid
C;C \mid
\kif\,B\,C\,C
$$
Here $B$ is a heap-independent boolean expression,
such as $x{=}0$. Commands $C$ are from the loop-free
simple imperative language. They can call
 module operations $k$, and manipulate heap cells;
command $[x]{:=}E$ assigns
$E$ to the heap cell $x$, and this assigned value is read
by $\klet\,y{=}[x]\,\kin\,C$, which also binds
$y$ to the read value and runs $C$ under this binding.

Properties of commands $C$ are specified using Hoare triples
$\Gamma \vdash \{\varphi\}C\{\psi\}$, where the context $\Gamma$
is a set of triples for module operations.
Figure~\ref{fig:proofrules} shows rules for proving
these properties. In the figure, we omit contexts, if the same
context $\Gamma$ is used for all the triples.

\begin{figure}[t]
\hrule
$$
\begin{array}{@{}c@{}}
\infer{
\{\varphi'\}C\{\psi'\}
}{
\lcheck(\varphi',\varphi)
\;\;\,
\varphi' \,{\models^1}\, \varphi
\;\;\,
\{\varphi\}C\{\psi\}
\;\;\,
\psi \,{\models^1}\, \psi'
\;\;\,
\lcheck(\psi,\psi')
}
\\[1ex]
\infer{
\{\varphi*\psi\}C\{\varphi'*\psi\}
}{
\{\varphi\}C\{\varphi'\}
}
\quad
\infer[x\,{\not\in}\,\FV(C)]{
\{\exists x.\,\varphi\}C\{\exists x.\,\psi\}
}{
\{\varphi\}C\{\psi\}
}
\\[1ex]
\infer{
\Gamma,\{\varphi\}k\{\psi\} \vdash \{\varphi\}k\{\psi\}
}{
}
\quad
\infer{
\{E\ipsto\blank\}[E]{:=}F\{E\ipsto F\}
}{}
\\[1ex]
\infer[x\,{\not\in}\,\FV(\psi)]{
\{\exists x. \varphi \,{*}\, E\ipsto x\}\klet\,x{=}[E]\,\kin\,C\{\psi\}
}{
\{\varphi \,{*}\, E\ipsto x\}C\{\psi\}
}
\\[1ex]
\infer{
\{\varphi\}C;C'\{\psi\}
}{
\{\varphi\}C\{\varphi'\}
\;\;
\{\varphi'\}C'\{\psi\}
}
\;\;\;
\infer{
\{\varphi\}\kif\,B\,C\,C'\{\psi\}
}{
\{\varphi \wedge B\}C\{\psi\}
\;\;
\{\varphi \wedge \neg B\}C'\{\psi\}
}
\end{array}
$$
\hrule
\caption{Proof Rules}\label{fig:proofrules}
\end{figure}

The rule of consequence deserves attention. Note that the rule uses
semantic implications $\models^1$ in the standard unary
interpretation, thus allowing the use of existing theorem provers for
separation logic. The rule does not allow all semantic implications,
but only those that pass our algorithm $\lcheck$, so as to ensure that
the implications can lift to the relational level. Our algorithm
$\lcheck(\varphi,\psi)$ performs two checks, and returns $\true$ only
when both succeed. The first check is whether $\varphi$ and $\psi$ can
be transformed to simple assertions $\varphi'$ and $\psi'$, using only
the distribution of $*$ over $\exists x$ and $\vee$ and distributive
lattice laws for $\vee$ and $\wedge$. If this check succeeds and gives
$\varphi'$ and $\psi'$, the algorithm transforms $\varphi' \models^1
\psi'$ to a set of implications of the form
\eqref{eqn:implication-form} in
Section~\ref{section:liftingtheoremsandcompleteness}
(Lemma~\ref{lemma:implication-reduction}). Then, for each implication
in the resulting set, it tests if any of the the three criteria for
lifting are met and returns true if that is always the
case\footnote{Recall that the failure of the lifting theorems do not
imply that a concrete implication cannot be lifted; consider, e.g.,
Example \ref{example:fan} and replace $1 \ipsto \_$ with $\true$
everywhere. One can sidestep the general lifting theorems and (try to)
verify directly the Parametricity Condition from
Definition~\ref{def:parametricitycondition} for all environments
$\eta$. It is, however, a semantic condition and probably undecidable
in general.}.

Commands $C$ are interpreted in a standard way, as functions of the
type: $\sem{C}_{\eta,u} \in \heap \rightarrow (\heap \cup
\{\wrong\})$. Here $\wrong$ denotes a memory error, and $\eta$ and $u$
are environments that provide the meanings of, respectively, free
ordinary variables and module operations. For instance,
$\sem{k}_{\eta,u}$ is $u(k)$.

Our semantics of triples, on the other hand, is not standard,
and uses the binary interpretation of assertions:
$(\eta,\rho,\vec{u}) \models^2 \{\varphi\}C\{\psi\}$ iff
$$
\begin{array}{@{}l@{}}
\forall r \in \irel_2.\;\forall f,g \in \heap.\;\;
(f,g) \,{\in}\, \sembin{\varphi}_{\eta,\rho} * r \implies {}
\\
\qquad\qquad\qquad
(\sem{C}_{\eta,u_1}(f),\sem{C}_{\eta,u_2}(g)) \in \sembin{\psi}_{\eta,\rho} * r.
\end{array}
$$
The environment $\rho$ provides the meanings of assertion variables,
and the $2$-dimensional vector $\vec{u}$ gives the two
meanings for module operations; intuitively, each $u_i$ corresponds
to the $i$-th module implementation. The interpretation means that
if two module implementations $\vec{u}$ are used by the same client $C$,
then these combinations should result in the same computation, in the sense
that they map $\varphi$-related input heaps to $\psi$-related outputs.
The satisfaction of triples can be extended to $(\eta,\rho,\vec{u})
\models^2 \Gamma$, by asking that
all triples in $\Gamma$ should hold wrt.\ $(\eta,\rho,\vec{u})$.
Using these satisfaction relations on triples and contexts, we define
the notion of $2$-validity of judgements: $\Gamma \vdash \{\varphi\}C\{\psi\}$ is $2$-valid iff
$$
\forall (\eta,\rho,\vec{u}).\;
(\eta,\rho,\vec{u}) \models^2 \Gamma
\implies
(\eta,\rho,\vec{u}) \models^2 \{\varphi\}C\{\psi\}.
$$
\begin{thm}\label{thm:parametricity-logic}
Every derivable $\Gamma \,{\vdash}\, \{\varphi\}C\{\psi\}$ is $2$-valid.
\end{thm}
It is this theorem that we use to derive the representation
independence results mentioned in the introduction. Consider again the
example in Figure~\ref{fig:counter}. Since the proof of the left hand
side client $C$ is derivable using the above rules in the context
\begin{align*}
  \Gamma = \{1{\ipsto}\blank\}&\init\{a\},\{a\}\incr\{a\}, \{a\}\gonext\{b\},\\
  \{b\}&\decr\{b\},\{b\}\final\{1{\ipsto}\blank\},
\end{align*}
we get $2$-validity of $\Gamma \vdash \{1\ipsto \blank\} C \{1\ipsto
\blank\}$. Instantiating, in the definition of 2-validity, $\rho$ with
the given coupling relations and $\vec{u}$ with the module
implementations gives us 
\[
  (\emptyset,\rho,\vec{u}) \models^2 \{1\ipsto \blank\} C \{1\ipsto \blank\},
\]
since we already know that the different operations respect the
coupling relations. Therefore, when we run the client $C$ with the
related module implementations, we find that $C$ maps $\sem{1\ipsto
\blank}^2$-related heaps (i.e, heaps with the same value at cell $1$)
to $\sem{1\ipsto \blank}^2$-related heaps again.


\section{Conclusion and Discussion} \label{section:conclusionanddiscussion}

In this paper, we have given a sound and, in a certain sense, complete
characterization of when \emph{semantic} implications in separation
logic with assertion variables can be lifted to a relational
interpretation. This characterization has, then, been used to identify
proofs of clients that respect the abstraction of module internals,
specified by means of assertion variables, and to show representation
independence for clients with such proofs. We hope that our results
provide a solid semantic basis for recent logic-based approaches to
data abstraction.

\paperskip{
We have demonstrated how our completeness results are of practical
import, by showing how bad clients not satisfying representation
independence can be constructed from implications not fulfilling the
conditions of our lifting theorem.
}

In earlier work, Banerjee and Naumann~\cite{banerjee:naumann:jacm}
studied relational parametricity for dynamically allocated heap
objects in a Java-like language. Later they extended their results to
cover clients programs that are correct with respect to specifications
following the ``Boogie methodology'' implemented in the Spec\#
verifier \cite{Banerjee-ECOOP05,Boogie}. In both works, Banerjee and
Naumann made use of a non-trivial semantic notion of confinement to
describe internal resources of a module; here instead we use
separation logic with assertions variables to describe which resources
are internal to the module.

Data abstraction and information hiding have been studied in logics
and specification languages for pointer programs, other than separation
logic. Good example projects are ESC-Modular-3~\cite{ESC-Modular3}, ESC-Java~\cite{ESC-Java} and Spec\#~\cite{SpecSharp},
some of which use concepts analogous to abstract predicates, called
abstract variables \cite{Leino-TOPLAS02}. It would be an interesting future direction to 
revisit the questions raised in the paper in the context of these
logics and specification languages.

Relational interpretations have also been used to give models of
programming languages with local state, which can validate
representation independence
results~\cite{OHearn:Tennent:95,Pitts:Stark:93,Benton:Leperchey:05,Dreyer:popl09}.
These results typically rely on the module allocating the private
state, whereas we use the power of separation logic and allow the
ownership transfer of states from client to module. For instance,
in the two-stage counter in the introduction, the ownership of the
cell $1$ is transferred from the client to the module upon calling
$\init$.  Even with this ownership transfer, representation
independence is guaranteed, because we consider only those clients
having (good) proofs in separation logic. This contrasts with
representation independence results in local state models, which
consider not some but all well-typed clients. The work by Banerjee and
Naumann~\cite{banerjee:naumann:jacm} discussed above also permits
ownership transfer.

\section*{Acknowledgement}
We would like to thank the anonymous referees for thorough
reviews and insightful comments; their work made this work better.
We are also grateful to John Wickerson and Peter O'Hearn for
their helpful comments. Yang was supported by EPSRC.


{
\bibliographystyle{abbrv}
\bibliography{bib}

\begin{thebibliography}{10}

\bibitem{Dreyer:popl09}
A.~Ahmed, D.~Dreyer, and A.~Rossberg.
\newblock State-dependent representation independence.
\newblock In {\em POPL}, pages 340--353, 2009.

\bibitem{banerjee:naumann:jacm}
A.~Banerjee and D.~Naumann.
\newblock Ownership confinement ensures representation independence for
  object-oriented programs.
\newblock {\em JACM}, 52(6), 2005.

\bibitem{Banerjee-ECOOP05}
A.~Banerjee and D.~A. Naumann.
\newblock State based ownership, reentrance, and encapsulation.
\newblock In {\em ECOOP}, pages 387--411, 2005.

\bibitem{Boogie}
M.~Barnett, B.-Y.~E. Chang, R.~DeLine, B.~Jacobs, and K.~R.~M. Leino.
\newblock Boogie: A modular reusable verifier for object-oriented programs.
\newblock In {\em FMCO}, pages 364--387, 2005.

\bibitem{SpecSharp}
M.~Barnett, R.~DeLine, M.~F{\"a}hndrich, B.~Jacobs, K.~R.~M. Leino, W.~Schulte,
  and H.~Venter.
\newblock The spec\# programming system: Challenges and directions.
\newblock In {\em VSTTE}, pages 144--152, 2005.

\bibitem{Benton:Leperchey:05}
N.~Benton and B.~Leperchey.
\newblock Relational reasoning in a nominal semantics for storage.
\newblock In {\em TLCA}, pages 86--101, 2005.

\bibitem{biering-birkedal-torpsmith-esop05}
B.~Biering, L.~Birkedal, and N.~{Torp-Smith}.
\newblock {BI} hyperdoctrines and higher-order separation logic.
\newblock In {\em ESOP}, 2005.

\bibitem{Biering:Birkedal:Torp-Smith:07}
B.~Biering, L.~Birkedal, and N.~Torp-Smith.
\newblock {BI}-hyperdoctrines, higher-order separation logic, and abstraction.
\newblock {\em TOPLAS}, 29(5), 2007.

\bibitem{Birkedal:Torp-Smith:Yang:05}
L.~Birkedal, N.~Torp-Smith, and H.~Yang.
\newblock Semantics of separation-logic typing and higher-order frame rules.
\newblock In {\em LICS}, pages 260--269, 2005.

\bibitem{Birkedal:Yang:07}
L.~Birkedal and H.~Yang.
\newblock Relational parametricity and separation logic.
\newblock In {\em FOSSACS}, pages 93--107, 2007.

\bibitem{DBLP:conf/concur/Brookes04}
S.~D. Brookes.
\newblock A semantics for concurrent separation logic.
\newblock In P.~Gardner and N.~Yoshida, editors, {\em CONCUR}, volume 3170 of
  {\em Lecture Notes in Computer Science}, pages 16--34. Springer, 2004.

\bibitem{Chin-popl08}
W.-N. Chin, C.~David, H.~H. Nguyen, and S.~Qin.
\newblock Enhancing modular oo verification with separation logic.
\newblock In {\em POPL}, 2008.

\bibitem{Distefano-TACAS06}
D.~Distefano, P.~W. O'Hearn, and H.~Yang.
\newblock A local shape analysis based on separation logic.
\newblock In {\em TACAS}, pages 287--302, 2006.

\bibitem{DistefanoJStar08}
D.~Distefano and M.~Parkinson.
\newblock {jStar}: towards practical verification for java.
\newblock In {\em OOPSLA}, pages 213--226, 2008.

\bibitem{distefano2008jstar}
D.~Distefano and M.~Parkinson~J.
\newblock jstar: Towards practical verification for java.
\newblock {\em ACM Sigplan Notices}, 43(10):213--226, 2008.

\bibitem{Dockins-aplas09}
R.~Dockins, A.~Hobor, and A.~Appel.
\newblock A fresh look at separation algebras and share accounting.
\newblock In {\em APLAS}, pages 161--177, 2009.

\bibitem{ESC-Java}
C.~Flanagan, K.~R.~M. Leino, M.~Lillibridge, G.~Nelson, J.~B. Saxe, and
  R.~Stata.
\newblock Extended static checking for java.
\newblock In {\em PLDI}, 2002.

\bibitem{JacobsVeriFast08}
B.~Jacobs and F.~Piessens.
\newblock The {VeriFast} program verifier.
\newblock Technical Report CW-520, Katholieke Universiteit Leuven, 2008.

\bibitem{LeeYangRasmussemCAV11}
O.~Lee, H.~Yang, and R.~Petersen.
\newblock Program analysis for overlaid data structures.
\newblock In {\em CAV}, 2011.

\bibitem{ESC-Modular3}
K.~R.~M. Leino and G.~Nelson.
\newblock An extended static checker for modular-3.
\newblock In {\em CC}, pages 302--305, 1998.

\bibitem{Leino-TOPLAS02}
K.~R.~M. Leino and G.~Nelson.
\newblock Data abstraction and information hiding.
\newblock {\em ACM Trans. Program. Lang. Syst.}, 24(5):491--553, 2002.

\bibitem{nanevski-htt-07}
A.~Nanevski, A.~Ahmed, G.~Morrisett, and L.~Birkedal.
\newblock Abstract predicates and mutable {ADTs} in {Hoare} type theory.
\newblock In {\em ESOP}, 2007.

\bibitem{BirkedalL:ynot-conf}
A.~Nanevski, G.~Morrisett, A.~Shinnar, P.~Govereau, and L.~Birkedal.
\newblock Ynot: dependent types for imperative programs.
\newblock In {\em ICFP}, pages 229--240, 2008.

\bibitem{OHearn:Tennent:95}
P.~O'Hearn and R.~Tennent.
\newblock Parametricity and local variables.
\newblock {\em JACM}, 42(3):658--709, May 1995.

\bibitem{DBLP:journals/toplas/OHearnYR09}
P.~W. O'Hearn, H.~Yang, and J.~C. Reynolds.
\newblock Separation and information hiding.
\newblock {\em ACM Trans. Program. Lang. Syst.}, 31(3), 2009.

\bibitem{Parkinson:Bierman:05}
M.~Parkinson and G.~Bierman.
\newblock Separation logic and abstraction.
\newblock In {\em POPL}, pages 247--258, 2005.

\bibitem{Parkinson:Bierman:08}
M.~Parkinson and G.~Bierman.
\newblock Separation logic, abstraction and inheritance.
\newblock In {\em POPL}, pages 75--86, 2008.

\bibitem{Pitts:Stark:93}
A.~Pitts and I.~Stark.
\newblock Observable properties of higher order functions that dynamically
  create local names, or: What's new?
\newblock In {\em MFCS}, pages 122--141, 1993.

\bibitem{Plotkin:Abadi:93}
G.~Plotkin and M.~Abadi.
\newblock A logic for parametric polymorphism.
\newblock In {\em TLCA}, pages 361--375, 1993.

\bibitem{Reynolds:83}
J.~C. Reynolds.
\newblock Types, abstraction, and parametric polymorphism.
\newblock In {\em Information Processing 83}, pages 513--523, 1983.

\bibitem{Strachey:67}
C.~Strachey.
\newblock Fundamental concepts in programming languages.
\newblock In {\em Proceedings of the 1967 International Summer School in
  Computer Programming, Copenhagen, Denmark.}, 1967.

\end{thebibliography}
}

\appendix
\section{Proofs of Lemma \ref{lem:diag-preservation} and Lemma \ref{lem:diag-semantics}}

\paragraph{\bf Lemma \ref{lem:diag-preservation}}
{\it
Function $\diag_n$ preserves the complete lattice structure
and the $*$ operator.
}
\proof
From the definition, it is immediate that $\diag_n(\heap) = \heap^n$ and
$\diag_n(\emptyset) = \emptyset$, the former because we have $[] \in \heap$.
Now consider a non-empty family $\{p_i\}_{i \in I}$ of predicates in
$\irel_1$. In order to show the preservation of the complete lattice
structure, we need to prove that
$$
\diag_n(\bigcap_{i\in I} p_i) = \bigcap_{i\in I} \diag_n(p_i)
\;\;\wedge\;\;
\bigcup_{i\in I} \diag_n(p_i) = \diag_n(\bigcup_{i\in I} p_i).
$$
The $\subseteq$ direction in both cases is easy; it follows
from the monotonicity of $\diag_n$.

We start with the $\supseteq$ direction for the meet operator.
Pick $(h_1,\ldots,h_n)$ from $\bigcap_{i\in I} \diag_n(p_i)$.
Then,
$$
\forall i \in I.\;\; (h_1,\ldots,h_n) \in \diag_n(p_i).
$$
By the definition of $\diag_n$, this means that
\begin{equation}
\label{eqn:diag-preservation1}
\forall i \in I.\;\;\exists f_i \in p_i.\;\;
f_i \sqsubseteq h_1
\;\wedge\;
\ldots
\;\wedge\;
f_i \sqsubseteq h_n.
\end{equation}
Let $f = \sum_{i \in I} f_i$. The sum here is well-defined, because
(a) there are only finitely many
$f$'s such that $f \sqsubseteq h_k$ for all $1 \leq k \leq n$,
and (b) any two such $f$ and $g$ should have the same value
for every location in $\dom(f) \cap \dom(g)$. Since all $f_i$'s
satisfy \eqref{eqn:diag-preservation1}, their sum $f$ also satisfies
$$
f \sqsubseteq h_1
\;\wedge\;
\ldots
\;\wedge\;
f \sqsubseteq h_n.
$$
Furthermore, $f \in \bigcap_{i \in I} p_i$, because $p_i$'s are upward
closed and $f$ is an extension of $f_i$ in $p_i$.
Hence, $\diag_n(\bigcap_{i \in I} p_i) \subseteq
\bigcap_{i\in I} \diag_n(p_i)$.

Next we prove the $\supseteq$ direction for
the join operator. Pick $(h_1,\ldots,h_n)$ from $\diag_n(\bigcup_{i\in I} p_i)$. Then,
$$
\exists i \in I.\;\;
\exists f \in p_i.\;\;
f \sqsubseteq h_1
\;\wedge\;
\ldots
\;\wedge\;
f \sqsubseteq h_n.
$$
Hence, by the definition of $\diag_n$,
$$
(h_1,\ldots,h_n) \;\;\in\;\;
\diag_n(p_i) \;\;\subseteq\;\;
\bigcup_{i \in I} \diag_n(p_i),
$$
as desired.

Finally, it remains to show that $\diag_n$ preserves
the $*$ operator. Consider predicates $p,q \in \irel_1$.
We need to prove that
$$
  \diag_n(p * q) \;=\; \diag_n(p) * \diag_n(q).
$$
Choose an arbitrary $(h_1,\ldots,h_n)$ from $\diag_n(p * q)$. By the
definition of $\diag_n(p*q)$, it follows that
$$
\begin{array}{@{}l@{}}
\exists f \in p.\;\;
\exists g \in q.\;\;
(\dom(f) \cap \dom(g) = \emptyset) \;\wedge\; {}
\\[0.5ex]
\quad
f \sqsubseteq h_1
\;\wedge\;
\ldots
\;\wedge\;
f \sqsubseteq h_n
\;\wedge\;
g \sqsubseteq h_1
\;\wedge\;
\ldots
\;\wedge\;
g \sqsubseteq h_n.
\end{array}
$$
Now, define $f_i = f$ and $g_i = h_i - f$
for $i \in \{1,\ldots,n\}$. Then,
$$
\begin{array}{@{}l@{}}
(\forall i \in \{1,\ldots,n\}.\;f_i \cdot g_i = h_i)
\\[0.5ex]
{} \;\wedge\;
(f_1,\ldots,f_n) \in \diag_n(p)
\;\wedge\;
(g_1,\ldots,g_n) \in \diag_n(q).
\end{array}
$$
Hence, $(h_1,\ldots,h_n) \in \diag_n(p)*\diag_n(q)$.
This shows that $\diag_n(p * q) \subseteq \diag_n(p)*\diag_n(q)$.
For the other inclusion, suppose that
$$
(h_1,\ldots,h_n) \;\in\; \diag_n(p)*\diag_n(q).
$$
Then, by the definition of $*$,
$$
\begin{array}{@{}l@{}}
\exists (f_1,\ldots,f_n) \in \diag_n(p).\;\;
\exists (g_1,\ldots,g_n) \in \diag_n(q).\;\;
\\[0.5ex]
\qquad
(\forall i \in \{1,\ldots,n\}.\;f_i \cdot g_i = h_i).
\end{array}
$$
Since $(f_1,\ldots,f_n) \in \diag_n(p)$ and
$(g_1,\ldots,g_n) \in \diag_n(q)$, there are $f \in p$
and $g \in q$ such that
$$
f \sqsubseteq f_1
\;\wedge\;
\ldots
\;\wedge\;
f \sqsubseteq f_n
\;\wedge\;
g \sqsubseteq g_1
\;\wedge\;
\ldots
\;\wedge\;
g \sqsubseteq g_n.
$$
Furthermore, since $f_1$ and $g_1$ have disjoint domains,
their subheaps $f$ and $g$ must have disjoint domains as well.
Consequently, $f \cdot g$ is well defined, and it satisfies
$$
f \cdot g \in p * q
\;\wedge\;
(\forall i \in \{1,\ldots,n\}.\;
f \cdot g \,\sqsubseteq\, f_i \cdot g_i \,=\, h_i).
$$
This implies that $(h_1,\ldots,h_n) \in \diag_n(p*q)$, as desired.
\qed

\paragraph{\bf Lemma \ref{lem:diag-semantics}}
{\it
For all $\varphi$ and $\eta,\rho,\rho'$,
if $\diag_n(\rho(a)) = \rho'(a)$ for every $a \in \avar$, we
have that $\diag_n(\semun{\varphi}_{\eta,\rho}) = \semn{\varphi}_{\eta,\rho'}$.
}
\proof
We prove by induction on the structure of $\varphi$.
All the inductive cases and the cases of $\true$ and $\false$
follow from the preservation result
of Lemma \ref{lem:diag-preservation}. Thus, it is sufficient to
show the lemma when $\varphi \equiv a$ or $\varphi \equiv P$.
When $\varphi \equiv a$, the assumption of the lemma
implies that
$$
\diag_n(\semun{a}_{\eta,\rho})
\;=\;
\diag_n(\rho(a))
\;=\;
\rho'(a)
\;=\;
\semn{a}_{\eta,\rho'}.
$$
When $\varphi \equiv P$, we note that
$\diag_n \circ \diag_1 = \diag_n$, and conclude
that
$$
\diag_n(\semun{P}_{\eta,\rho})
\;=\;
\diag_n(\diag_1(\semprim{P}_\eta))
\;=\;
\diag_n(\semprim{P}_\eta)
\;=\;
\semn{P}_{\eta,\rho'}.
$$
\qed

\section{Proof of Lemma~\ref{lemma:implication-reduction}}
\label{appendix:proof-implication-reduction}

\paragraph{\bf Lemma \ref{lemma:implication-reduction}}
{\it There is an algorithm taking simple assertions $\varphi,\psi$ and
returning finitely many implications $\{\varphi^l {\implies}
\psi^l\}_{l \in L}$, such that (a) $\varphi^l {\implies} \psi^l$
has the form~\eqref{eqn:implication-form} and (b) for any $n \in
\{1,2\}$, we have that $\varphi \models^n \psi$ holds iff
$\varphi^l \models^n \psi^l$ holds for all $l \in L$.}

\proof
The algorithm first transforms $\psi$ in the conjunctive normal
form, using proof rules in classical logic, which hold in all the
$n$-ary semantics.  This gives an implication of the form:
$$
\bigvee_{i{=}1}^I
\bigwedge_{j{=}1}^J
\varphi_{(i,j)} * \vec{a}_{(i,j)}
\implies
\bigwedge_{k{=}1}^K
\bigvee_{l{=}1}^L
\psi_{(k,l)} * \vec{b}_{(k,l)}.
$$
Then, the algorithm constructs the below set:
$$
\left\{
\bigwedge_{j{=}1}^J
\varphi_{(i,j)} * \vec{a}_{(i,j)}
\implies
\bigvee_{l{=}1}^L
\psi_{(k,l)} * \vec{b}_{(k,l)}
\right\}_{1\leq i \leq I, 1 \leq k \leq K}.
$$
Finally, it removes, in each implication, all the disjuncts that include
assertion variables not appearing on the LHS of the implication; if all
disjuncts are removed, $\false$ is the new RHS. The outcome of this removal
becomes the result of the algorithm.
\qed

\section{Layouts that Lift}
\begin{lem}[Segregation] \label{lemma:segregation}
For any $I, J\geq 1$ there are non-empty, finite
\emph{segregating} subsets $\Seg^{I,J}_{i,j} \subseteq \pinteger$
for all $1 \leq i \leq I$ and $1 \leq j \leq J$ with these
properties:
\begin{enumerate}[\em(1)]
 \item $\forall 1 \,{\leq}\, i_1, i_2 \,{\leq}\, I.\ \ \bigcup_{1 \leq j \leq J}\Seg^{I,J}_{i_1, j} = \bigcup_{1 \leq j \leq J} \Seg^{I,J}_{i_2,j}.$
 \item $\forall 1 \,{\leq}\, i \,{\leq}\, I.\ \
        \forall 1 \,{\leq}\, j_1 \,{\neq}\, j_2 \,{\leq}\, J.\ \  \Seg^{I,J}_{i,j_1} \cap \Seg^{I,J}_{i,j_2}  = \emptyset.$
 \item $\forall 1 \,{\leq}\, i_1 \,{\neq}\, i_2 \,{\leq}\, I.\ \
        \forall 1 \,{\leq}\, j_1, j_2  \,{\leq}\, J.\ \ \Seg^{I,J}_{i_1,j_1} \cap \Seg^{I,J}_{i_2, j_2} \neq  \emptyset.$
\end{enumerate}
By 1 we define $\Seg^{I,J} = \bigcup_{1 \leq j \leq
J}\Seg^{I,J}_{i, j}$ for any $1 \leq i \leq I$.
\end{lem}

\paragraph{\bf Theorem \ref{theorem:shadow} (Shadow-Lift)}
{\it Unary $\eta$-validity of an implication implies the PC if each
dashed line has a label that is not a label on a solid line and
each disjunct has an occurrence of a variable that is not a label
on a solid line. Spelling it out in symbols, we require, with $L =
\{(i,j) \mid 1\leq i \leq M \wedge 1 \leq j \leq N\}$, that
$$
\begin{array}{@{}l@{}}
 \forall (i,j) \in L.\ \; \Pi(i) \ngeq \Omega(j) \implies {}
\\
  \;\;\; \exists c \in V.\ \; \Pi(i)(c) < \Omega(j)(c) \,\wedge\, {}
\\
  \;\;\;\;\; \bigl(\forall (k,l) \in L.\ \; \Pi(k) \geq \Omega(l) \implies  \Pi(k)(c) = \Omega(l)(c)\bigr)
\end{array}
$$
and
$$
\begin{array}{@{}l@{}}
   \forall 1 \leq j \leq N.\ \; \exists c \in V.\ \;
   \Omega(j)(c) > 0\,\wedge\, {}
\\
   \;\;\;
   \bigl(\forall (k,l) \in L.\ \; \Pi(k) \geq \Omega(l) \implies
   \Pi(k)(c) = \Omega(l)(c)\bigr).
\end{array}
$$
}
\proof
Assume that we have an implication of the form
\eqref{eqn:implication-form} in
Section~\ref{section:liftingtheoremsandcompleteness} and an
appropriate environment $\eta$, that the stated criterion on the
variable layout holds and that we have unary $\eta$-validity. We
must show that the PC holds.

According to Definition \ref{def:parametricitycondition} we assume
that we have heaps $h, h_1, \ldots, h_M \in \heap$ with $h_i
\sqsubseteq h$ and $h_i \in \semun{\varphi_i}_\eta$ for all $1
\leq i \leq M$. The core of the proof is the construction of a
particular environment $\rho: V \to \irel_1$. For that purpose we
need some notation. For a subset $M \subseteq \pinteger$ we denote
by $[M]$ the heap that has domain $M$ and stores some fixed value,
say $0$, at all these locations. Let $C \subseteq V$ be the set of
assertion variables that do not occur as labels on solid edges,
i.e., for a $c \in V$ we have that $c \in C$ iff
$$
\begin{array}{l}
\forall 1 \leq i \leq M.\;\; \forall 1 \leq j \leq N.\;\;
\\[1ex]
\qquad \Pi(i) \geq \Omega(j) \implies \Pi(i)(c) = \Omega(j)(c).
\end{array}
$$
For each $1 \leq i \leq M$ we let $K_i$ be the set of second
indices of all variables in conjunct $i$ that lie in $C$, i.e., we
set $K_i = \{1 \leq k \leq M_i \mid a_{i,k} \in C\}$. If
non-empty, we let $k_i = \min(K_i)$.

We now define $\rho(c) = \heap$ for $c \in V \setminus C$. For a
variable $c \in C$ we let $\rho(c)$ be the union of
\[
  \bigcup_{\scriptsize\begin{array}{c}1 \leq i \leq M, \\K_i \neq \emptyset, \\a_{i,k_i} \equiv c\end{array}}
  \Big\{(h-h_i) \cdot [\Seg^{M,K}_{i,k_i} + L] \cdot \prod_{\scriptsize\begin{array}{c}1 \leq k \leq K, \\k \notin K_i\end{array}} [\Seg^{M,K}_{i,k} +   L]\Big\}^\uparrow\
\]
and
\[
  \bigcup_{1 \leq i \leq M, K_i \neq \emptyset, k \in K_i \setminus\{k_i\}, a_{i,k} \equiv c} [\Seg^{M,K}_{i,k} +L],
\]
where we have used $K = \max\{M_1, \ldots, M_M\}$ and $L =
\max(\dom(h)\cup\{0\})$. For each $1 \leq i \leq M$ we can write
$h\cdot[\Seg^{M,K}+L]$ as the following product
\[
   h_i \cdot (h-h_i) \cdot \prod_{k \in K_i} [\Seg^{M,K}_{i,k}+L] \cdot \prod_{1 \leq k \leq K, k \notin K_i} [\Seg^{M,K}_{i,k} +  L],
\]
which implies that we have $h\cdot[\Seg^{M,K}+L]$ a member of
$\semun{\varphi_i*a_{i,1}* \cdots * a_{i,M_i}}_{\eta, \rho}$. In
summary, we have shown that $h\cdot[\Seg^{M,K}+L]$ lies in the
unary interpretation of the left hand side in the environments
$\eta$ and $\rho$. By assumption, the same must hold for the right
hand side and from this we aim to derive the PC.

We now know that $h\cdot[\Seg^{M,K}+L]$ lies in the interpretation
of some disjunct, say disjunct $j$. This means that
\begin{align*}
    h\cdot[\Seg^{M,K}+L] & \in \semun{\psi_j * b_{j,1} * \cdots * b_{j,
    N_j}}_{\eta,\rho}\\
    & = \semun{\psi_j}_\eta * \prod_{k \in J}\rho(b_{j,k}),
\end{align*}
where $J = \{1 \leq k \leq N_j \mid b_{j,k} \in C\}$ is the set of
second indices of variables of disjunct $j$ that are in $C$. By
the second assumption of the theorem we know that $J \neq
\emptyset$. We write
 \[h\cdot[\Seg^{M,K}+L] = g \cdot \prod_{k \in J} g_k\]
for $g \in \semun{\psi_j}_\eta$ and $g_k \in \rho(b_{j,k})$ for
each $k \in J$. By the properties of segregating sets we get that
there must be a common $1 \leq i \leq M$ such that for all $k \in
J$ there is $l_k \in K_i$ with
  \[[\Seg^{M,K}_{i,l_k} + L] \sqsubseteq g_k, \]
i.e., the $g_k$'s are all `from the same conjunct'. But this
implies $\Pi(i)(c) \geq \Omega(j)(c)$ for all $c \in C$ as the
segregating sets are non-empty. But then $\Pi(i)(c) \geq
\Omega(j)(c)$ must hold for $c \in V\setminus C$ too by the first
assumption of the lemma and so $\Pi(i) \geq \Omega(j)$. Also we
must have $\Pi(i)(c) = \Omega(j)(c)$ for each $c \in C$ by
definition of $C$. By construction we have
\[
  \dom\left(\prod_{k \in J} g_k\right) \supseteq \dom(h-h_i) \cup
  (\Seg^{M,K} + L)
\]
But then $\dom(g) \subseteq h_i$ and so we have $h_i \in
\semun{\psi_j}_\eta$ too and we have proved the first option of
the PC.
\qed

\paragraph{\bf Theorem \ref{theorem:balloon} (Balloon-Lift)}
{\it Unary $\eta$-validity of an implication implies the PC if there is
a subset $B \subseteq V$ with the following three properties.
First, each conjunct has at most one occurrence of a variable from
$B$, i.e.,
$$
\forall 1\leq i\leq M.\ \sum_{c \in B} \Pi(i)(c) \leq 1.
$$
Second, each disjunct is empty or has exactly one occurrence of a
variable from $B$, i.e.,
$$
\forall 1\leq j\leq N.\ \sum_{c \in V} \Omega(j)(c) = 0 \,\vee\,
\sum_{c\in B} \Omega(j)(c) = 1.
$$
Third,  each dashed line must have a label from $B$. That is, when
$L = \{(i,j) \mid 1\leq i \leq M \wedge 1\leq j \leq N\}$,
$$
 \forall (i,j) \,{\in}\, L.
 \;
 \Pi(i) \ngeq \Omega(j) \implies \exists c \,{\in}\, B.\ \Pi(i)(c) < \Omega(j)(c).
$$
}
\proof
Assume that we have an implication of the form
\eqref{eqn:implication-form} in
Section~\ref{section:liftingtheoremsandcompleteness} and an
appropriate environment $\eta$, that the stated criterion on the
variable layout holds and that we have unary $\eta$-validity. We
must show that the PC holds.

According to Definition \ref{def:parametricitycondition} we assume
that we have heaps $h, h_1, \ldots, h_M \in \heap$ with $h_i
\sqsubseteq h$ and $h_i \in \semun{\varphi_i}_\eta$ for all $1
\leq i \leq M$. The core of the proof is the construction of a
particular environment $\rho: V \to \irel_1$. We define $\rho(c) =
\heap$ for $c \in V \setminus B$. For a variable $c \in B$ we let
$\rho(c)$ be the following union
\[
  \bigcup_{1 \leq i \leq M, 1 \leq k \leq M_i, a_{i,k} \equiv c}
  \{h-h_i\}^\uparrow.
\]
For each $1 \leq i \leq M$ we can write $h = h_i \cdot (h-h_i)$ and so we
have $h$ in $\semun{\varphi_i*a_{i,1}* \cdots * a_{i,M_i}}_{\eta, \rho}$ by
the first of the original assumptions on the set $B$. In summary, we have
shown that $h$ lies in the unary interpretation of the left hand side in the
environments $\eta$ and $\rho$. By assumption, the same must hold for the
right hand side and from this we aim to derive the PC.

We now know that $h$ lies in the interpretation of some disjunct,
say disjunct $j$. If this disjunct is empty we have proved the
second option of the PC. Otherwise we know that there is exactly
one $1 \leq k \leq N_j$ such that $b_{j,k} \in B$. But then we
have
\[
  h \in \semun{\psi_j * b_{j,1} * \cdots * b_{j, N_j}}_{\eta,\rho} = \semun{\psi_j}_\eta *
  \rho(b_{j,k}).
\]
We write
 \[h = g \cdot g_k\]
for $g \in \semun{\psi_j}_\eta$ and $g_k \in \rho(b_{j,k})$. There
must be an $1 \leq i \leq M$ such that $g_k \sqsupseteq h-h_i$ and
such that $\Pi(i)(b_{j,k}) = \Omega(j)(b_{j,k}) = 1$. The first
gives $h_i \in \semun{\psi_j}_\eta$ and the second implies $\Pi(i)
\geq \Omega(j)$ by the third assumption on $B$. And we have
arrived at the first option of the PC.
\qed

\section{Completeness} \label{appendix:completeness}
\begin{lem}[Fan-Counter] \label{lemma:fan}
Suppose that the layout of variables is as follows. There are at
least two conjuncts, i.e., $M \geq 2$, and one conjunct has the
property that each variable occurring in the conjunct also occurs
as a label of a solid line leaving the conjunct and ending in a
non-empty disjunct. In symbols the latter is
\begin{align*}
    &\exists 1 \leq i\leq M.\ \forall c \in V.\ \Pi(i)(c) > 0 \implies \\
    & \quad \exists 1 \leq j \leq N.\  \Pi(i) \geq \Omega(j) \wedge \Pi(i)(c) > \Omega(j)(c)\ \wedge\\
    & \quad \quad \exists d \in V.\ \Omega(j)(d) > 0.
\end{align*}
Then there are choices of $\varphi_i$'s and $\psi_j$'s with no
variables such that the implication holds on the unary level but
not on the binary level.
\end{lem}
In the search for counterexamples we may without loss of
generality assume the negation of the conditions of the above
lemma. This means, provided at least two conjuncts, that for any
non-empty set of solid lines leaving one common conjunct and
ending in non-empty disjuncts there is a variable that occurs in
the conjunct but is not a label of either of the lines. If,
loosely phrased, we invalidate that variable, then all the solid
lines break down, i.e., become dashed.

\begin{lem}[X-Counter] \label{lemma:x}
Suppose that the layout of variables is as follows. There are two
distinct conjuncts $i_0$ and $i_1$ and two distinct non-empty
disjuncts $j_0$ and $j_1$ such that $\Pi(i_0) \ngeq \Omega(j_0)$
while $\Pi(i_0) \geq \Omega(j_1)$, $\Pi(i_1) \geq \Omega(j_0)$ and
$\Pi(i_1) \geq \Omega(j_1)$. Then there are choices of
$\varphi_i$'s and $\psi_j$'s that the implication holds on the
unary level but not on the binary level.
\end{lem}
Again we may without loss of generality assume that the negation
of this lemma holds when building counterexamples. Picture the
graph of the implication without empty disjuncts and without
dashed lines. The negation of the above means that we may arrive
at all vertices in the connected component containing some vertex
by paths from that vertex of length 2 or less. Also all connected
components are complete, in particular no two vertices with a
dashed line between them can belong to the same component.

\begin{lem}[Bridge-Counter] \label{lemma:bridge}
Suppose that the layout of variables is as follows. There are at
least two conjuncts, i.e., $M \geq 2$, all disjuncts are non-empty
and there is a dashed line with labels that all occur as labels on
solid lines too. In symbols the last demand is
\begin{align*}
    & \exists 1 \leq i \leq M.\ \exists 1 \leq j \leq N.\ \Pi(i) \ngeq \Omega(j) \    \wedge \\
    & \quad \forall c \in V.\ \Pi(i)(c) < \Omega(j)(c) \implies \\
    & \quad \quad \exists 1 \leq k \leq M.\ \exists 1 \leq l \leq    N.\ \\
    & \quad \quad \quad\Pi(k) \geq \Omega(l) \wedge \Pi(k)(c) > \Omega(l)(c).
\end{align*}
Then there are choices of $\varphi_i$'s and $\psi_j$'s with no
variables such that the implication holds on the unary level but
not on the binary level.
\end{lem}
This lemma deals with the case of a variable layout with at least
two conjuncts and no empty disjuncts but where the first condition
of Theorem \ref{theorem:shadow} fails.

\begin{lem}[All-Out-Counter] \label{lemma:all-out}
Suppose that the layout of variables is as follows. There are at
least two conjuncts, i.e., $M \geq 2$, at least one non-empty
disjunct and for each variable one of the following two holds:
Either the variable occurs as a label on a solid line ending in a
non-empty disjunct. Or it occurs at least twice in a conjunct and
we have an empty disjunct. In symbols the variable condition is
\begin{align*}
    & \forall c \in V.\ \big(\exists 1 \leq i \leq M.\ \exists 1 \leq j \leq N.\ \Pi(i) \geq \Omega(j)\ \wedge \\
    & \quad\Pi(i)(c) > \Omega(j)(c) \wedge \exists d \in V.\ \Omega(j)(d) > 0\big) \ \vee \\
    & \quad \big(\exists 1 \leq i \leq M.\ \Pi(i)(c) \geq 2\ \wedge \\
    & \quad \quad \exists 1\leq j\leq N.\ \forall d \in V.\ \Omega(j)(d) =    0\big).
\end{align*}
Then there are choices of $\varphi_i$'s and $\psi_j$'s
$\varphi_i$'s and $\psi_j$'s with no variables  such that the
implication holds on the unary level but not on the binary level.
\end{lem}
This lemma deals with two cases. The first is the case of a
variable layout with at least two conjuncts and no empty disjuncts
but where the second condition of Theorem \ref{theorem:shadow}
fails while the first holds. The second is the case of a variable
layout with at least two conjuncts, at least one empty disjunct
and no dashed lines for which Theorem \ref{theorem:balloon} fails.

\section{Higher Arities and Parametricity} \label{appendix:higheraritiesandparametricity}

\paragraph{\bf Theorem~\ref{theorem:nbigimpliesgyc}}{\it For an implication of the form \eqref{eqn:implication-form}  and
an appropriate environment $\eta$ we have that $n$-ary $\eta$- validity
implies the PC if $n \geq \max\{2,M_1, \ldots, M_M\}$.}

\proof
Assume that we have an implication of the form \eqref{eqn:implication-form}
in Section~\ref{section:liftingtheoremsandcompleteness} and an appropriate
environment $\eta$, that $n \geq \max\{2,M_1, \ldots, M_M\}$ and that we have
$n$-ary $\eta$-validity. We must show that the PC holds.

According to Definition \ref{def:parametricitycondition} we assume
that we have heaps $h, h_1, \ldots, h_M \in \heap$ with $h_i
\sqsubseteq h$ and $h_i \in \semun{\varphi_i}_\eta$ for all $1
\leq i \leq M$. The core of the proof is the construction of a
particular environment $\rho: V \to \irel_n$. For that purpose we
need some notation. Define, for each $1 \leq k \leq n$, a map
$\gamma_k: \heap \to \heap^n$ by letting
$$
  \gamma_k(h) = \big(\overbrace{[],\ldots,[]}^{k-1},h,\overbrace{[],\ldots,[]}^{n-k}\big)
$$
for any $h \in \heap$, i.e., it returns the $n$-tuple that has $h$
as the $k$-th entry and the empty heap everywhere else. Similarly, we
define $\delta: \heap \to \heap^n$ by setting
$$
  \delta(h) = \big(\overbrace{h,\ldots,h}^n\big)
$$
for any $h \in \heap$, i.e., it returns the $n$-tuple that has $h$
as all entries. For a subset $M \subseteq \pinteger$ we denote by
$[M]$ the heap that has domain $M$ and stores some fixed value,
say $0$, at all these locations.

For a variable $c \in V$ we now define $\rho(c)$ to be the
following union of relations in $\irel_n$:
  \[\bigcup_{1\leq i \leq M, 1 \leq k \leq M_i, a_{(i,k)} \equiv c}
  \{\gamma_k(h-h_i) \cdot \gamma_1([\Seg^{M,K}_{i,k}+L])\}^\uparrow\]
where $K = \max\{M_1, \ldots, M_M\}$, $L = \max(\dom(h))$. This is
well-defined because of our assumption that $n \geq \max\{M_1,
\ldots, M_M\}$. For each $1 \leq i \leq M$ we have that
\begin{align*}
  \delta(h) & = \delta(h_i) \cdot \gamma_1(h-h_i) \cdot \cdots \cdot \gamma_n(h-h_i) \\
  & \sqsupseteq \delta(h_i) \cdot \gamma_1(h-h_i) \cdot \cdots \cdot \gamma_{M_i}(h-h_i)
\end{align*}
where we use the extension order for heap tuples defined by
pointwise extension in all entries.
Also, we have that
\begin{align*}
  [\Seg^{M,K}+L] & = [\Seg^{M,K}_{i,1}+L] \cdot \cdots \cdot [\Seg^{M,K}_{i,K}+L] \\
  & \sqsupseteq [\Seg^{M,K}_{i,1}+L] \cdot \cdots \cdot [\Seg^{M,K}_{i,M_i}+L].
\end{align*}
This gives us that $\delta(h) \cdot \gamma_1([\Seg^{M,K}+L])$
extends the following $n$-tuple of heaps:
  \[\delta(h_i) \cdot \prod_{1 \leq k \leq M_i} \gamma_k(h-h_i) \cdot
  \gamma_1([\Seg^{M,K}_{i,k}+L])\]
which again means that $\delta(h) \cdot \gamma_1([\Seg^{M,K}+L])$
lies in
  \[\semn{\varphi_i*a_{i,1}* \cdots * a_{i,M_i}}_{\eta, \rho}.\]
In summary, we have shown that $\delta(h) \cdot
\gamma_1([\Seg^{M,K}+L])$ lies in the $n$-ary interpretation of
the left hand side in the environments $\eta$ and $\rho$. By
assumption, the same must hold for the right hand side and from
this we aim to derive the PC.

There is $1 \leq j \leq N$ such that we have
  \[\delta(h) \cdot \gamma_1([\Seg^{M,K}+L]) \in \semn{\psi_j*b_{j,1}* \cdots * b_{j,N_j}}_{\eta, \rho}.\]
Consider first the case of a non-empty disjunct, i.e., the case
$N_j > 0$. We split along the disjunct and get
  \[\delta(h) \cdot \gamma_1([\Seg^{M,K}+L]) = \delta(g) \cdot \mathbf{g}_1 \cdot \cdots \cdot \mathbf{g}_{N_j}\]
for $g \in \semun{\psi_j}_\eta$ and $\mathbf{g}_k \in \rho(b_{j,k})$ for all
$1 \leq k \leq N_j$. By the properties of segregating sets we get that there
must be a common $1 \leq i \leq M$ such that for all $1 \leq k \leq N_j$
there is $1 \leq k_k \leq M_i$ with
  \[\gamma_{k_k}(h-h_i) \cdot \gamma_1([\Seg^{M,K}_{i,k_k}+L]) \sqsubseteq \mathbf{g}_k, \]
i.e., the $\mathbf{g}_k$'s are all `from the same conjunct'. But
this implies $\Pi(i) \geq \Omega(j)$ as the segregating sets are
non-empty. Also the above equality enforces $\dom(g) \subseteq
\dom(h)$ by the definition of $\gamma_1$. Indeed we must have
$\dom(g) \subseteq \dom(h_i)$ since in particular we have
  \[\gamma_{1_k}(h-h_i) \cdot \gamma_1([\Seg^{M,K}_{i,1_k}+L]) \sqsubseteq \mathbf{g}_1. \]
But then $g \sqsubseteq h_i$ so we have $h_i \in
\semun{\psi_j}_\eta$ too and the first option of the PC holds.

We consider now the case of an empty disjunct, i.e., the case $N_j
= 0$. As above we split along the disjunct and get
  \[\delta(h) \cdot \gamma_1([\Seg^{M,K}+L]) = \delta(g) \cdot \mathbf{g}\]
for $g \in \semun{\psi_j}_\eta$ and $\mathbf{g} \in \heap^n$.
Again we must have $\dom(g) \subseteq \dom(h)$ which implies $g
\sqsubseteq h$ and the second option of the PC holds.
\qed

\section{Proof of Theorem~\ref{thm:parametricity-logic}}

\paragraph{\bf Theorem \ref{thm:parametricity-logic}}{\it Every
derivable $\Gamma \,{\vdash}\, \{\varphi\}C\{\psi\}$ is $2$-valid.}
\proof
We will show that all the rules in Figure~\ref{fig:proofrules}
are sound. This lets us prove the theorem by induction on
the height of the derivation of a judgment,
because using the soundness of the rules, we can handle
all the base and inductive cases.

Let's start with the rule for the module operation $k$.
Suppose that we have $(\eta,\rho,\vec{u}) \models^2
(\Gamma, \{\varphi\}k\{\psi\})$. Then, by the definition of
$\models^2$, we should have
$(\eta,\rho,\vec{u}) \models^2 \{\varphi\}k\{\psi\}$
as well. From this follows the soundness of the rule.

Next, consider four rules: (a) the frame rule for adding $-*\varphi$
to the pre and post-conditions, (b) the rule for adding $\exists
x$ to the pre and post-conditions, (c) the rule for sequencing,
and (d) the rule for the conditional statement.
All these rules are sound, because of the following four facts:
$$
\begin{array}{@{}l@{}}
(\eta,\rho,\vec{u})
\models^2 \{\varphi\}C\{\varphi'\}
\\
\qquad
{} \implies
(\eta,\rho,\vec{u})
\models^2 \{\varphi * \psi\}C\{\varphi' * \psi\}
\\
\\
(x \not\in \FV(C))
\;\wedge\;
(\eta,\rho,\vec{u})
\models^2 \{\varphi\}C\{\psi\}
\\
\qquad
{} \implies
(\eta,\rho,\vec{u})
\models^2 \{\exists x.\varphi\}C\{\exists x.\psi\}
\\
\\
(\eta,\rho,\vec{u})
\models^2 \{\varphi\}C\{\varphi'\}
\;\wedge\;
(\eta,\rho,\vec{u})
\models^2 \{\varphi'\}C'\{\psi\}
\\
\qquad
{} \implies
(\eta,\rho,\vec{u})
\models^2 \{\varphi\}C;C'\{\psi\}
\\
\\
(\eta,\rho,\vec{u})
\models^2 \{\varphi \wedge B\}C\{\psi\}
\;\wedge\; {}
\\
(\eta,\rho,\vec{u})
\models^2 \{\varphi \wedge \neg B\}C'\{\psi\}
\\
\qquad
{} \implies
(\eta,\rho,\vec{u})
\models^2 \{\varphi\}\kif\,B\,C\,C'\{\psi\}
\end{array}
$$
The first fact is an easy consequence of using
the quantification over $\irel_2$ in the semantics of triples.
The second also follows easily from the semantics of triples and the fact that
$\sem{C}_{\eta,u_i} =
\sem{C}_{\eta[x{\mapsto}v],u_i}$ for all $v \in \integer$,
as long as one remembers that the $*$ operator distributes over
union. The third and fourth are not different, and they follow
from the semantics of triples and commands. Here we will go through the details
of proving the fourth fact. Consider $(\eta,\rho,\vec{u})$ satisfying
the assumption of the fact. Pick $r \in \irel_2$ and heaps $f,g$ such that
$$
  (f,g) \;\in\;\sembin{\varphi}_{\eta,\rho} * r.
$$
Now we do the case analysis on whether $\sem{B}_\eta$ is true or not.
If it is true, then
$$
\begin{array}{@{}l@{}}
  (f,g) \;\in\;\sembin{\varphi \wedge B}_{\eta,\rho} * r
\\[1ex]
  {} \;\;\wedge\;\;
  \sem{\kif\,B\,C\,C'}_{\eta,u_1}(f) \,=\, \sem{C}_{\eta,u_1}(f)
\\[1ex]
  {} \;\;\wedge\;\;
  \sem{\kif\,B\,C\,C'}_{\eta,u_2}(g) \,=\, \sem{C}_{\eta,u_2}(g).
\end{array}
$$
Hence, by assumption, we get that
$$
\begin{array}{@{}l@{}}
  \bigl(\,
   \sem{\kif\,B\,C\,C'}_{\eta,u_1}(f),\;
   \sem{\kif\,B\,C\,C'}_{\eta,u_2}(g)
  \,\bigr)
\\[1ex]
   \qquad\qquad
   {} \;\;=\;\;
   \bigl(\,
   \sem{C}_{\eta,u_1}(f),\;
   \sem{C}_{\eta,u_2}(g)
   \,\bigr)
  \;\;\in\;\;
  \sembin{\psi}_{\eta,\rho} * r.
\end{array}
$$
If $\sem{B}_\eta$ is not true, we reason similarly, but with
$C'$ instead of $C$, and get that
$$
  \bigl(\,
   \sem{\kif\,B\,C\,C'}_{\eta,u_1}(f),\;
   \sem{\kif\,B\,C\,C'}_{\eta,u_2}(g)
  \,\bigr)
  \;\;\in\;\;
  \sembin{\psi}_{\eta,\rho} * r.
$$
We have just shown that in both cases,
the outcomes of the conditional
statements are related by $\sembin{\psi}_{\eta,\rho} * r$, as claimed
by the fourth fact.

We move on to the rules for heap update and dereference.
They are sound because of the below two facts:
$$
\begin{array}{@{}l@{}}
(\eta,\rho,\vec{u})
\models^2
\{E\ipsto\blank\}[E]{:=}F\{E\ipsto F\}
\\
\\
x \not\in \FV(\psi)
\;\;\wedge\;\;
(\eta,\rho,\vec{u})
\models^2
\{\varphi \,{*}\, E\ipsto x\}C\{\psi\}
\\
\quad
{}\implies
(\eta,\rho,\vec{u})
\models^2
\{\exists x. \varphi \,{*}\, E\ipsto x\}\klet\,x{=}[E]\,\kin\,C\{\psi\}
\end{array}
$$
To prove the first, we pick $(\eta,\rho,\vec{u})$,
a relation $r \in \irel_2$ and heaps $f,g$ such that
$$
  (f,g) \;\in\;\sembin{E\,\ipsto\,\blank}_{\eta,\rho} * r.
$$
Then, there exist heaps $h,f_1,g_1$ such that
$$
  (f,g) = (h,h) \cdot (f_1,g_1) \;\;\wedge\;\;
  \sem{E}_\eta \in \dom(h) \;\;\wedge\;\;
  (f_1,g_1) \in r.
$$
Thus,
$$
\begin{array}{@{}l@{}}
(\,
\sem{[E]{:=}F}_{\eta,u_1}(f),\;\,
\sem{[E]{:=}F}_{\eta,u_2}(g)
\,)
\\[1ex]
\qquad
{} \;=\;
\bigl(\,
 h[\sem{E}_\eta{\mapsto}\sem{F}_\eta] \cdot f_1,\;\,
 h[\sem{E}_\eta{\mapsto}\sem{F}_\eta] \cdot g_1
\,\bigr)
\\[1ex]
\qquad
{} \;\in\;
\sembin{E\ipsto F}_{\eta,\rho} \,*\, r,
\end{array}
$$
as desired by the first fact. For the proof of the second fact,
suppose that the assumption of the second fact holds,
and pick $r \in \irel_2$ and heaps $f,g$ such that
$$
  (f,g) \;\in\;\sembin{\exists x.\,\varphi * E\ipsto x}_{\eta,\rho} * r.
$$
Then, there exists an integer $v$ and heaps $h,f_1,g_1, f_2, g_2$ such that
$$
\begin{array}{@{}l@{}}
  (f,g) = (h,h) \cdot (f_1,g_1) \cdot (f_2, g_2)\;\;\wedge\;\;
  h \in \semun{E\ipsto x}_{\eta[x{\mapsto}v]}  
\\[1ex]
  \qquad
  {} \;\;\wedge\;\;
  (f_1,g_1) \in \sembin{\varphi}_{\eta[x{\mapsto}v],\rho} \;\;\wedge\;\;
  (f_2,g_2) \in r    
\end{array}
$$
Thus, $(f,g) \in \sembin{\varphi * E\ipsto x}_{\eta[x{\mapsto}v],\rho} \,*\,
r$, and $f(\sem{E}_\eta) = g(\sem{E}_\eta) = v$. Using these and the
assumed triple of the second fact, we derive
the below:
$$
\begin{array}{@{}l@{}}
(\,
\sem{\klet\,x{=}[E]\,\kin\,C}_{\eta,u_1}(f),\;\,
\sem{\klet\,x{=}[E]\,\kin\,C}_{\eta,u_2}(g)
\,)
\\[1ex]
\qquad
{} \;=\;
\bigl(\,
 \sem{C}_{\eta[x{\mapsto}v],u_1}(f),\;\,
 \sem{C}_{\eta[x{\mapsto}v],u_2}(g)
\,\bigr)
\\[1ex]
\qquad
{} \;\in\;
\sembin{\psi}_{\eta[x{\mapsto}v],\rho} * r
\\[1ex]
\qquad
{} \;=\;
\sembin{\psi}_{\eta,\rho} * r.
\end{array}
$$
The last equality holds, because $x$ does not appear in $\varphi$.
We have just proved that the output states of two dereferencing
commands are $(\sembin{\psi}_{\eta,\rho} * r)$-related, as claimed
by the second fact.

Finally, we prove that the rule of consequence is sound.
It is sufficient to show that
$$
\begin{array}{@{}l@{}}
\lcheck(\varphi',\varphi)
\;\;\wedge\;\;
\varphi' \models^1 \varphi
\;\;\wedge\;\; {}
\\[1ex]
\lcheck(\psi,\psi')
\;\;\wedge\;\;
\psi \models^1 \psi'
\;\;\wedge\;\;
(\eta,\rho,\vec{u}) \models^2 \{\varphi\}C\{\psi\}
\\[1ex]
{}
\;\;\implies\;\;
(\eta,\rho,\vec{u}) \models^2 \{\varphi'\}C\{\psi'\}.
\end{array}
$$
From the first four conjuncts of the assumption, it follows
that
$$
\varphi' \models^2 \varphi
\;\;\;\;\wedge\;\;\;\;
\psi \models^2 \psi'.
$$
This is due to the correctness of $\lcheck$,
which holds because all the transformations used in the
first check of $\lcheck$ are based on semantic
equivalences holding in $\irel_2$ and
the second lifting check is sound because of our
lifting theorems. In order to prove the conclusion
of the above implication,
 pick $r \in \irel_2$ and heaps $f,g$ such that
$$
  (f,g) \;\in\;\sembin{\varphi'}_{\eta,\rho} * r.
$$
Since the $*$ operator is monotone and
$\varphi' \models^2 \varphi$, we get that
$$
  (f,g) \;\in\;\sembin{\varphi}_{\eta,\rho} * r.
$$
This relationship and the assumed triple $\{\varphi\}C\{\psi\}$,
then, imply the below:
$$
  (\,\sem{C}_{\eta,u_1}(f),\,\sem{C}_{\eta,u_2}(g)\,)
  \;\in\;\sembin{\psi}_{\eta,\rho} * r.
$$
Again, since $\psi \models^2 \psi'$, the monotonicity of the $*$
operator implies that
$$
  (\,\sem{C}_{\eta,u_1}(f),\,\sem{C}_{\eta,u_2}(g)\,)
  \;\in\;\sembin{\psi'}_{\eta,\rho} * r.
$$
Note that this is the conclusion that we are looking for.
\qed

\end{document}